\documentclass{aastex631}

\graphicspath{{./}{figures/}}
\shortauthors{Dalal et al.}
\usepackage{xcolor}
\usepackage{soul}

\begin{document}

\title{Differential behaviors of suprathermal $^4$He and Fe populations in the interplanetary medium during solar cycle 24} 

\author{Bijoy Dalal}
\affiliation{Physical Research Laboratory, Ahmedabad - 380009, India}
\affiliation{Indian Institute of Technology Gandhinagar, Gandhinagar - 382055, India}

\author{D. Chakrabarty}
\affiliation{Physical Research Laboratory, Ahmedabad - 380009, India}

\author{N. Srivastava}
\affiliation{Udaipur Solar Observatory, Physical Research Laboratory, 
Udaipur - 313001, India}

\begin{abstract}

Investigations on the solar cycle variation of the properties of suprathermal populations (H and other heavy ions like $^4$He, $^3$He, C, O and Fe) in the solar wind are sparse and hence, poorly understood. In the present investigation, solar cycle variations of ``quiet'' time suprathermal elements are investigated using $<$ $\sim$ 1 MeV/n particle flux data obtained from Ultra-Low Energy Isotope Spectrometer on board Advanced Composition Explorer satellite during the solar cycle 23 and 24. The analysis reveals that helium ($^4$He) shows zero or positive lags with respect to sunspot numbers in solar cycle 23 while it shows zero or negative lag in solar cycle 24. On the contrary, although iron (Fe) shows zero or positive lag in cycle 23 similar to $^4$He, it shows only zero lag in cycle 24 and no negative lag is seen. Further, significant differences in the spectral indices are seen between $^4$He and Fe in cycle 24 compared to the cycle 23. These results suggest that generation mechanisms responsible for suprathermal $^4$He and Fe underwent changes in cycle 24 and these mechanisms are probably dependent on the first ionization potential and mass to charge ratio. This proposition gets credence from the fact that changes in the lag and spectral slopes for C and O are not significantly different in cycle 23 and 24.

\end{abstract}


\section{Introduction} \label{sec:intro}

Suprathermal particles with energies from $\sim$ 10 KeV per nucleon (keV/n) to $\sim$ 1 MeV per nucleon (MeV/n) are thought to act as seed populations for further acceleration by interplanetary (IP) shocks associated with solar eruptive events like coronal mass ejections (CMEs) (\citealp{Gosling_et_al_1981, Desai_et_al_2003, Desai_et_al_2004} etc.) and co-rotating interaction regions (CIRs) (e.g., \citealp{Fisk_and_Lee_1980, Chotoo_et_al_2000, Allen_et_al_2019}). The two most widely known acceleration mechanisms namely first order Fermi acceleration (or diffusive shock acceleration) (\citealp{Krymskii_1977, Bell_1978} etc.) and the second order Fermi acceleration \citep{Fermi_1949} necessitate the initial presence of suprathermal particles at the acceleration framework. Energetic protons as well as heavy ions from $^4$He to Fe and beyond constitute suprathermal ion pool in the IP medium. Compositional abundance studies reveal that possible sources of suprathermal ion pool include solar wind ions \citep{Desai_et_al_2003}, particles associated with previously occurred transient events (\citealp{Fisk_and_Lee_1980, Giacalone_et_al_2002} etc.), interstellar pick-up ions \citep{Allen_et_al_2019}. In general, dominant contribution from the pick-up ions in the suprathermal populations is observed beyond 1 AU \citep{Fisk_1976}. Suprathermal particles exhibit power law distribution, also known as ``quiet'' time tail, when the velocity distribution function (differential directional flux) is plotted against velocity (energy). Often, spectral index of -5 (-1.5) \citep{Fisk_and_Gloeckler_2006, Fisk_and_Gloeckler_2007} has been reported in the past regardless of the species considered.  In this work, spectral index in differential flux vs energy convention (-1.5) is chosen. There had been a few reports in the past (e.g. \citealp{Gloeckler_2003, Mason_et_al_2012, Dayeh_et_al_2017}) that showed that this spectral index deviates substantially from the -1.5 value.  The reasons for significant variations in the spectral index is a topic that is still poorly understood. Further, it is also not abundantly clear why different elements in the suprathermal ion pool should follow similar spectral index \citep{Fisk_and_Gloeckler_2006, Mason_et_al_2008}. These issues are important and need critical attention as these might throw light on the generation of suprathermal ion population in the IP medium.
Enhancements of suprathermal protons and other heavy ions during CIRs within and beyond 1 AU have been reported by many authors (\citealp{Mason_et_al_1997, Mason_et_al_2008, Mason_et_al_2012, Chotoo_et_al_2000, Bucik_et_al_2009, Ebert_et_al_2012, Filwett_et_al_2017, Yu_et_al_2018, Allen_et_al_2019, Allen_et_al_2020} etc.). As a possible source of energization of suprathermal particles associated with CIRs, many have suggested diffusive shock acceleration by the forward and reverse shocks bounding the compression region of CIRs (e.g. \citealp{Palmer_and_Gosling_1978, Fisk_and_Lee_1980}). \cite{Fisk_and_Lee_1980} proposed a model in which particles accelerated by shock fonts can be decelerated to lower energies by the effect of adiabatic expansion and associated cooling  in the IP medium. However, the spectral rollover predicted by this model below 0.5 MeV/n could not be observed \citep{Mason_et_al_1997}. The gradual compression region bounded by the shock pair associated with a CIR is also considered as an effective source of acceleration under suitable conditions \citep{Mason_2000, Giacalone_et_al_2002}, although the shock pairs might not be adequately formed within 1 AU on many occasions. Nevertheless, presence of suprathermal populations in the IP medium in absence of such compression region and shock structures makes it difficult to comprehend the acceleration processes involved. Further, there are also evidences that ``quiet'' time suprathermal particles depend on solar activity. By studying the relative abundances of ``quiet'' time suprathermal ions at 1 AU, \cite{Kecskemety_et_al_2011} found out that during minimum of solar cycle, suprathermal Fe/O ratio resembles the corresponding ratio in the solar wind. These results are supported later on by \cite{Dayeh_et_al_2017}, in which solar cycle dependence of suprathermal C/O, Fe/O and $^3$He/$^4$He with very strong correlations with the yearly averaged sunspot numbers (SSNs) were reported. They argued that suprathermal particles are transported from remote places during the ``quiet'' times and are accelerated locally.
Regardless of the acceleration process involved, it is expected that during the course of transport of the suprathermal particles through the IP medium, the particles may show systematic time delay with solar activity proxies (for example, SSNs) and this may provide important clues for understanding the source of these particle populations. Although indicated in a few earlier works (e.g. \citealp{Mason_et_al_2012, Allen_et_al_2019}) detailed investigations on these time delays are sparse. Further, comparison of these time delays and spectral indices for various elements for multiple solar cycles may lead to new insights related to the role of solar/IP processes for the generation of these particles. Keeping these issues in mind, the suprathermal particle flux data from the Ultra Low Energy Isotope Spectrometer (ULEIS) \citep{Mason_et_al_1998} on board the Advanced Composition Explorer (ACE) \citep{Stone_et_al_1998} for solar cycle 23 and 24 (henceforth, SC23 and SC24 respectively) have been extensively analyzed and the results are presented in the present work. The results reveal important differences between SC23 and SC24 as far as the above aspects are concerned.

\section{Dataset}
Advanced Composition Explorer or ACE is a spin-stabilized spacecraft revolving around the Lagrangian point (L1) of the Sun-Earth system in a halo orbit. In this work, one-hour-integrated differential directional flux (level 2) data corresponding to different energy channels for H, $^4$He, $^3$He, O, C and Fe from Mar 01, 1998 to Aug 31, 2020 obtained from the Ultra-Low Energy Isotope Spectrometer (ULEIS) on board ACE are used. ULEIS is a time of flight mass spectrometer, which measures the time of flight ($\tau$) and deposited energy (E) of isotopes with Z= 2 to 28. Using the measured $\tau$ and corrected energy (see \citealp{Mason_et_al_1998} for details), mass (M) of the isotope is determined. The uncertainty in the energy measurement in ULEIS may affect mass separation near the low energy threshold in presence of significant noise (see \citealp{Mason_et_al_1998} and references cited therein). This problem is particularly relevant for $^3$He and $^4$He. Nevertheless, we have avoided using low energy ($<$ 100 keV/n) channels for all the elements in our investigation. The dataset spans almost 22 years covering the SC23 and SC24 and is available at \url{https://cdaweb.gsfc.nasa.gov/cgi-bin/eval2.cgi}. A list of co-rotating Interaction Region (CIR) events, identified based on the measurements at the L1 point and reported in \cite{Allen_et_al_2019}, is also used to remove the concomitant enhancements in the suprathermal flux associated with the CIR events. In addition, as stated in the next section, we also remove any other transient event that causes enhancements in the suprathermal flux. We have termed these events as “non-CIR” transient events. This step leads to a time series corresponding to ``quiet'' time. Here ``quiet'' is put inside the inverted commas as we will argue later that in either case – whether it is exhaustive removal of the transient events or removal of transient events based on a cut-off flux level - the propagation of suprathermal flux from the transient events (CIR, ICME) into the ``quiet'' time suprathermal ion pool cannot be completely eliminated in presence of significant lag with the sunspot numbers (SSN). We will come back to this point later on in more detail. The daily averaged sunspot number (SSN) data are available at \url{http://www.sidc.be/silso/datafiles}.

\begin{figure}[ht!]
\plotone{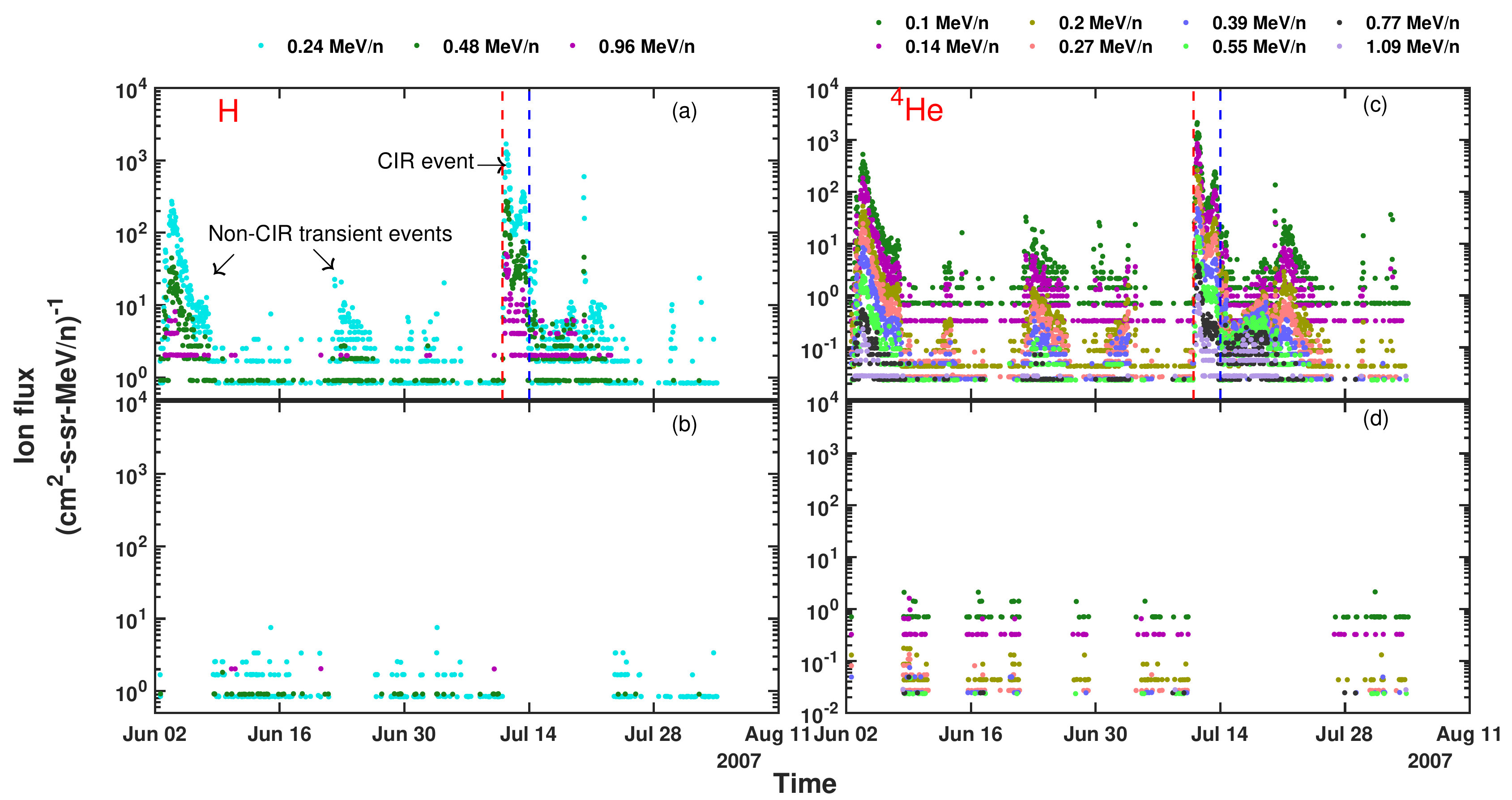}
\caption{The left column shows the proton flux data (a) before and (b) after the removal of the transient events. The dashed vertical lines indicate the start (red) and end (blue) times of the CIR event during this period as observed by ULEIS/ACE. The non-CIR transient events are indicated by arrows in (a). Different energy channels are represented by the mean of the lower and upper energy limits and are shown by colored dots. The right column shows the $^4$He flux variation before [(c)] and after [(d)] the removal of transient events similar to what has been done for the proton channels.\label{fig:fig1}}
\end{figure}

\section{Selection of quiet periods, sensitivity analyses, and validation}
As indicated in the previous section, the dataset used in the present work pertain to periods devoid of transient events (e.g. CIR and any other non-CIR events) that cause enhancements in the suprathermal flux. From the entire dataset spanning over 22 years, it is noted that transient CIR or non-CIR events take the flux levels much higher than the background flux level encountered at other times. Therefore, these transient flux variations are apparently easy to remove by choosing a threshold flux level (e.g. \citealp{Kecskemety_et_al_2011}). However, in the present work, we do not initially choose any cut-off flux level as a reference level to remove the transient events. Rather, we remove the flux variations associated with these transient variations in totality (from base flux level to peak flux level) so that the propagation of the peak flux into the ``quiet'' time flux is minimized significantly. It is possible that some minor transient injections will remain in the time series even after doing this. Therefore, we also perform detailed sensitivity analyses and cross-check critically to show that the results obtained through present analysis procedure (no cut-off level, removal of transient events) remains nearly unchanged with those obtained with four different cut-off flux levels (much lower than the peak levels of the transient injection events from base to peak level). Note, at least two of these cut-off levels are below the lowest flux level encountered during solar minima (between SC23 and SC24) in the original time series. We also subject these sensitivity analyses for different averaging windows (in days). These detailed sensitivity analyses with different cut-off levels and averaging windows are provided as supplementary Figures \ref{fig:S1}-\ref{fig:S12}. It is found that the correlation coefficients and the lags, thus derived, between suprathermal fluxes for each element (at different energy channels) and SSN remain nearly invariant for different choices of cut-off levels and averaging windows. Therefore, we will not discuss this point subsequently in this paper.
\par
\begin{figure}[ht!]
\plotone{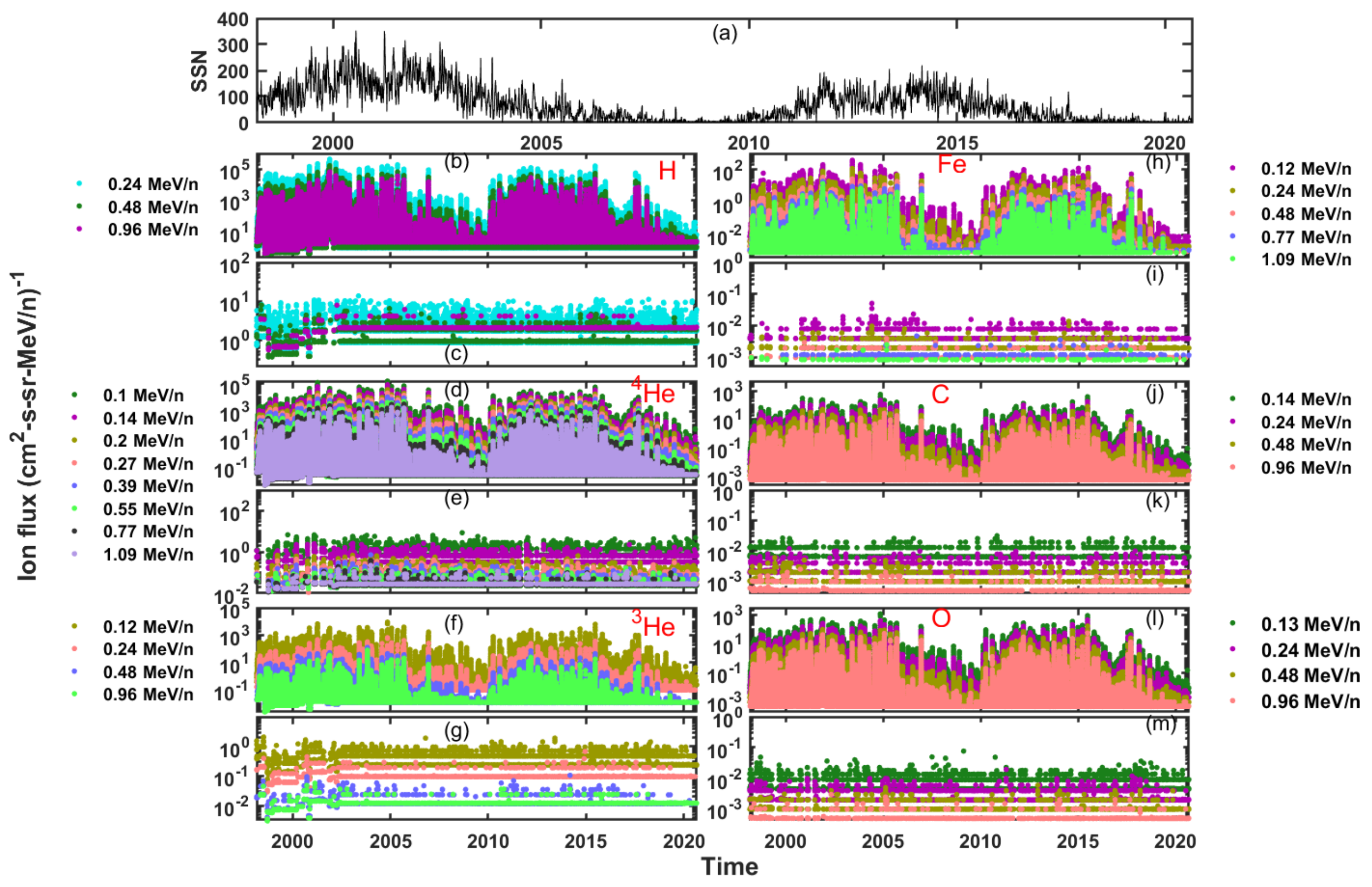}
\caption{Original vis-à-vis modified (after removal of the transient events) variations of the fluxes of different elements for the entire period (1998-2020) along with the SSN data used in this paper (panel `a'). Subplots (b), (d), (f), (h), (j), and (l) show the original flux variations of H, $^4$He, $^3$He, Fe, C, and O respectively. The modified (``quiet'') fluxes of the elements in the similar order are plotted in subplots (c), (e), (g), (i), (k), and (m). Legends for different energy channels of an element are marked by colored dots and are appropriately placed at the left and right of the plots\label{fig:fig2}}
\end{figure}
\begin{figure}[ht!]
\plotone{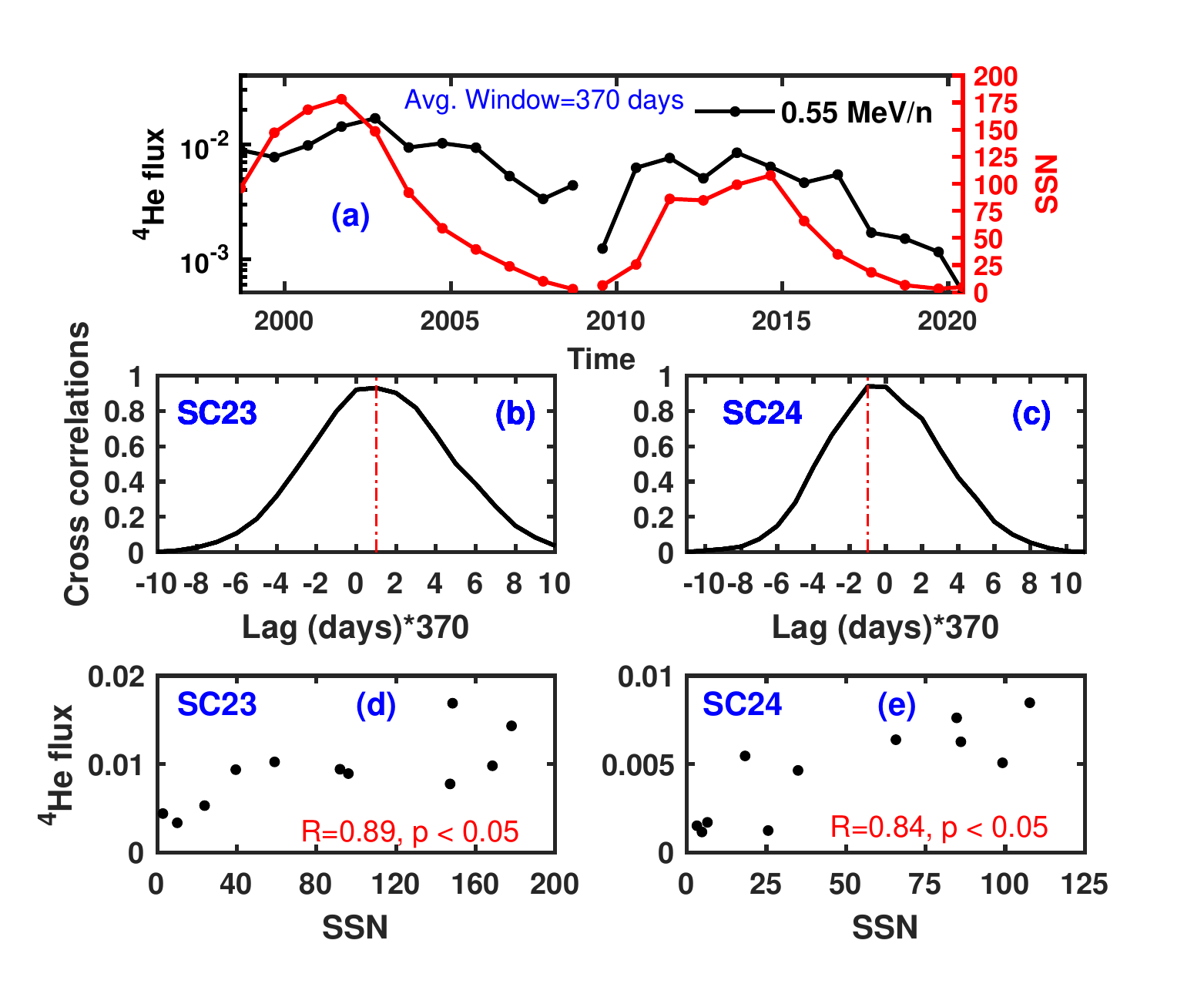}
\caption{Panel (a) shows typical variations of 370 days averaged $^4$He flux (black) corresponding to 0.55 MeV/n (mentioned at the right corner)  and SSN (red). Variations of normalized cross correlation coefficients between $^4$He flux and SSN with respect to lags (in multiples of 370 days) for SC23 and SC24 are shown in subplot (b) and (c) respectively. Scatter plots in subplots (d) and (e) represent variations of averaged and adjusted $^4$He fluxes with respect to SSN in SC23 and SC24 respectively (see text for details). Pearson correlation coefficients between flux and SSN and corresponding p values are mentioned at the bottom of panel (d) and (e).\label{fig:fig3}}
\end{figure}

We now come back to the methodology adopted in this work. This aspect is illustrated in Figure \ref{fig:fig1}. Figure \ref{fig:fig1}(a) and \ref{fig:fig1}(b) show the variations in the proton fluxes for the original and ``quiet'' (after removal of the transient events from base to peak) dataset respectively from 02 June, 2007 to 03 August, 2007. Different colors correspond to different energy channels. The mean of the upper and lower limits of each energy channel is taken to represent (legend) the corresponding energy channel. During the representative interval shown in Figure \ref{fig:fig1}, there was one CIR event (according to \citealp{Allen_et_al_2019}), the start and end times of which are marked in red and blue vertical dashed line respectively. The other transient events are termed as ‘non-CIR transient events’ and no attempt is made here to characterize those events. Figure \ref{fig:fig1} (c) and \ref{fig:fig1}(d) show the original and corresponding ``quiet'' time fluxes of $^4$He at various energy channels during the same time interval as that of H. It is to be noted that durations of transient events removed from time series of $^4$He may be different from those removed from H time series. This is because of the fact that the enhancement and depletion of different elements at the measurement location are not simultaneous in many cases \citep{Reames_2018}. Considering this important aspect, we have removed the transient events for each element manually. The complete exclusion of the transient events for the whole duration of the data set for each element under consideration is shown in Figure \ref{fig:fig2}. The daily averaged SSN data are shown in Figure \ref{fig:fig2}(a). Please note that detailed correlation analyses of this SSN time series are performed with the elemental flux time series using different averaging windows. These aspects are discussed in Figure \ref{fig:fig3}-\ref{fig:fig5} and in Figure \ref{fig:S1}-\ref{fig:S12}. In Figure \ref{fig:fig2}, subplots (b), (d), (f), (h), (j) and, (l) show the temporal variation of suprathermal H, $^4$He, $^3$He, Fe, C and, O fluxes respectively. Corresponding ``quiet'' time fluxes are shown in subplots (c), (e), (g), (i), (k) and, (m) respectively. The mean values (up to two decimal place) of each energy channel for H, $^4$He and $^3$He are mentioned at the top left of Figure \ref{fig:fig2} using colored dots as legends. The same is done for Fe, C and O and mentioned at the top right of the figure. 

It is to be noted at this point that all the sub-plots of Figure \ref{fig:fig2} are significantly compressed in time as these contain data spanning over 22 years. Although there are apparent bunching of spikes in some of the ``quiet'' time series plots (for example, in (h)) giving an impression of flux changes associated with transient events, it is verified that those spikes are formed due to temporal compression of the plot and constituent data points are temporarily far apart from each other. Another important point to be noted here is that the resultant ``quiet'' flux variations for all the elements lie below conspicuous threshold levels that emerge out automatically for each element. ``quiet'' time particle flux data obtained through this method are then analyzed further and results are discussed in subsequent sections. 

\section{Results}
\subsection{Correlation coefficients and lags with the variations in the SSN in SC23 and SC24}

Correlation between ``quiet'' suprathermal fluxes and SSN during SC23 and SC24 is investigated in this section. Fluxes of different particles at different energy channels and SSN have been subjected to averaging over 240 days to 400 days in steps of 10 days. In each such step, maximum cross correlation coefficient and corresponding lag between flux and SSN is found out. In order to get an idea about the significance of these correlations, we use these lags, apply to the time series data and calculate the Pearson correlation coefficients (CC) and corresponding p values (probability of acceptance of null hypothesis that these correlations are occurring by chance) between lagged fluxes and SSN. The method adopted in this work to compute CCs and lags is depicted in Figure \ref{fig:fig3}. Figure \ref{fig:fig3}(a) shows typical variations of 370 days' averaged $^4$He flux with energy 0.55 MeV/n and SSN for SC23 and SC24. In each case, SSN variation is subjected to positive and negative lags with respect to the flux variation to maximize cross correlation coefficient. The results obtained by this method for SC23 and SC24 are shown in Figures \ref{fig:fig3}(b) and \ref{fig:fig3}(c) respectively. The lags corresponding to the maximum cross correlation coefficients (marked by vertical red dot-dash lines) are considered as the lags of interest. A positive lag value indicates that the flux variation lags the SSN variation. On the other hand, a negative lag reveals that the flux variation leads the SSN variation, which essentially means that increase in flux has started earlier than the increase in SSN. The above two scenarios are clear from Figure \ref{fig:fig3}(b) and \ref{fig:fig3}(c) respectively. It can be seen that 370 days' averaged $^4$He flux lags the SSN variation in SC23 by 370 days and leads it in SC24 by the same days. Once the lag is noted, flux variation is adjusted for the lag and Pearson CC and corresponding p value mentioned are estimated. Figure \ref{fig:fig3}(d) and \ref{fig:fig3}(e) show Pearson CCs and p values for SC23 and SC24 respectively. Note, in both the cases R is higher than 0.8 and p values are $<$ 0.05 that indicate that these high correlations are real.  

\begin{figure}[ht!]
\plotone{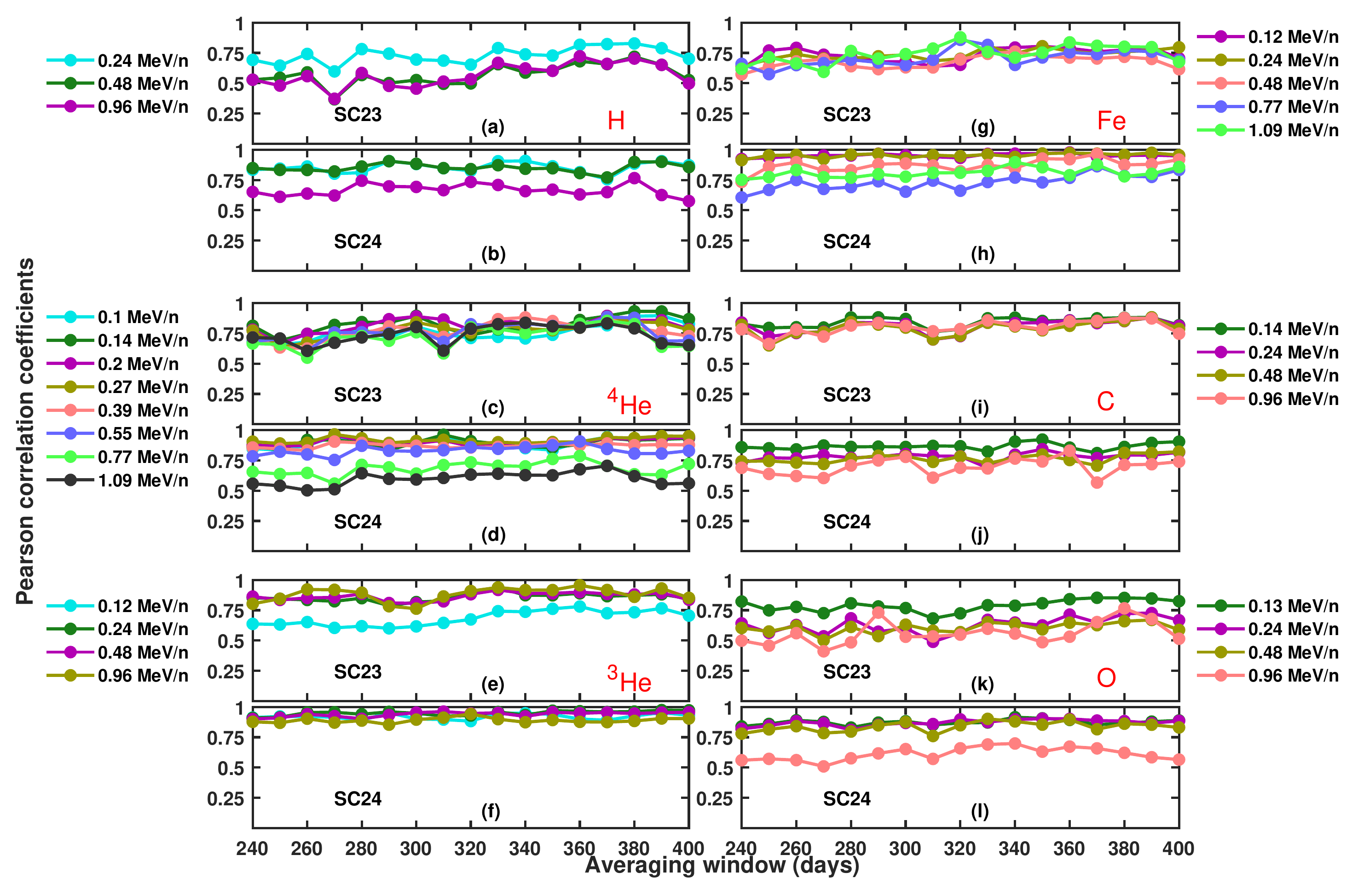}
\caption{Variations of Pearson’s correlation coefficients (CCs) between lagged fluxes and SSN with respect to averaging window. Panel (a), (c), (e), (g), (i), and (k) show the variations of CCs for H, $^4$He, $^3$He, Fe, C, and O respectively during SC23. Panel (b), (d), (f), (h), (j), and (l) represent the same for SC24. Energy channels corresponding to H, $^4$He, and $^3$He are written on the left side of the left column of the figure. The same corresponding to Fe, C, and O are mentioned on the right of the right column of the figure.\label{fig:fig4}}
\end{figure}

Figure \ref{fig:fig4} shows the variation of Pearson CCs with respect to the averaging window during SC23 and SC24 for various elements. Every vertical pair of subplots in this figure corresponds to the variations of Pearson CCs of an element in SC23 and SC24. It can be seen from Figure \ref{fig:fig4} that CC values corresponding to any particular energy channel of any element do not vary significantly depending on the averaging windows. In most of the cases it is observed that p$<$0.05. 

\begin{figure}[ht!]
\plotone{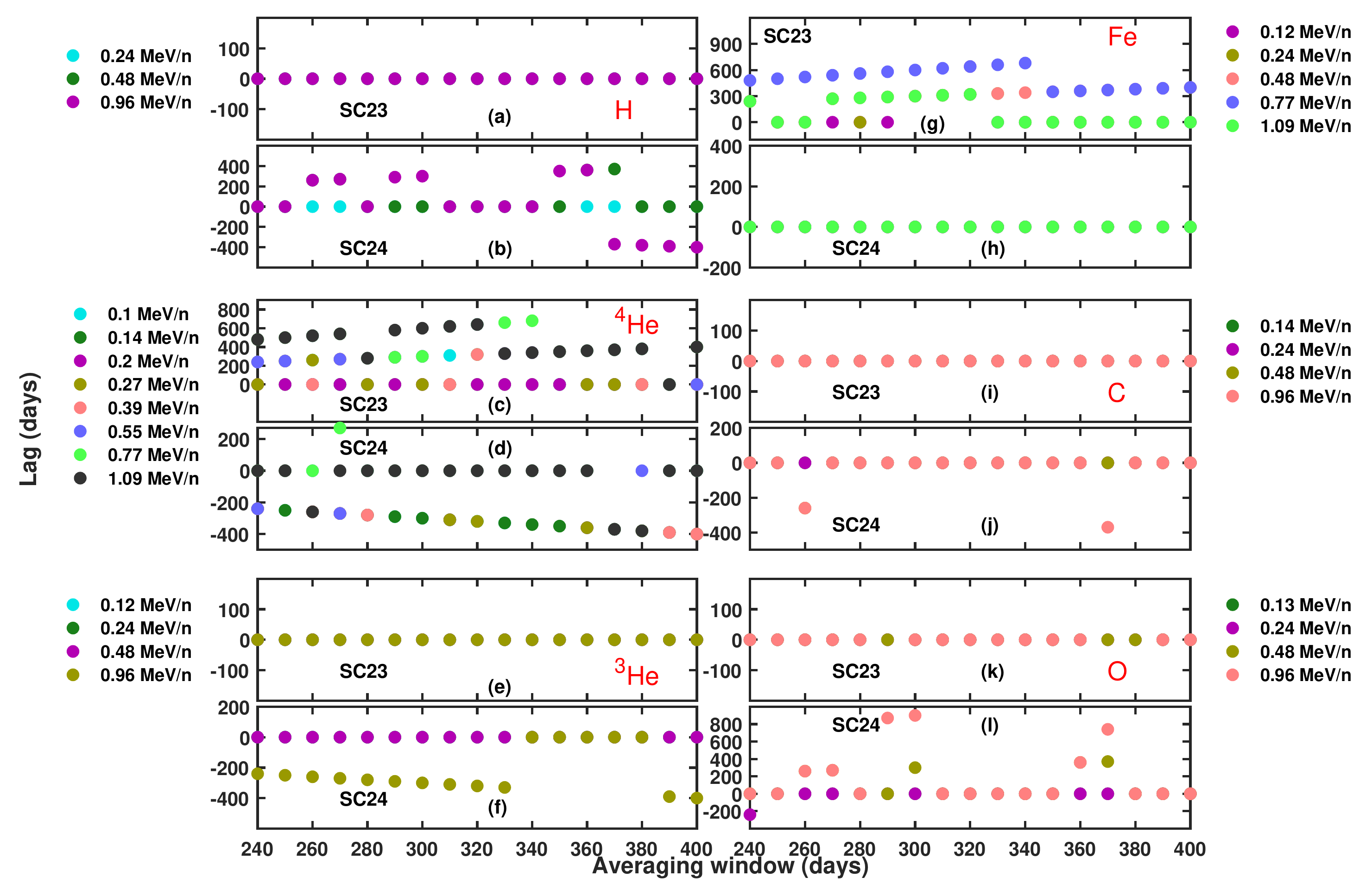}
\caption{Variations of lags between averaged fluxes and SSN with respect to averaging window. Panel (a), (c), (e), (g), (i), and (k) show the variations of lags for H, $^4$He, $^3$He, Fe, C, and O respectively during SC23. Panel (b), (d), (f), (h), (j), and (l) represent the same during SC24. Energy channels corresponding to H, $^4$He, and $^3$He are written on the left side of the left column of the figure. The same corresponding to Fe, C, and O are mentioned on the right of the right column of the figure.\label{fig:fig5}}
\end{figure}

Variations of lags corresponding to each point in Figure \ref{fig:fig4} are shown in Figure \ref{fig:fig5}. Some interesting outcomes from this figure are discussed in the following part of this section. ``quiet'' time suprathermal H at L1 point does not show any noticeable time delay with respect to SSN variation in SC23. The time lags are inconsistent in SC24 as positive, zero, and negative lags are observed in different energy channels of H. Interestingly, although $^4$He shows zero and positive lags in SC23, zero and negative lags are observed on some occasions in SC24.

$^3$He flux variation does not lag the SSN variation in both SC23 and SC24 except for at 0.96 MeV/n energy channel when negative lags are seen in SC24 for some averaging periods. While zero or positive lags are observed for Fe fluxes in SC23, predominantly zero lags are there in SC24. C and O show mostly zero lags in both the solar cycles.

\section{Variations in spectral index of different elements at different phases of SC23 and SC24}

\begin{figure}[ht!]
\plotone{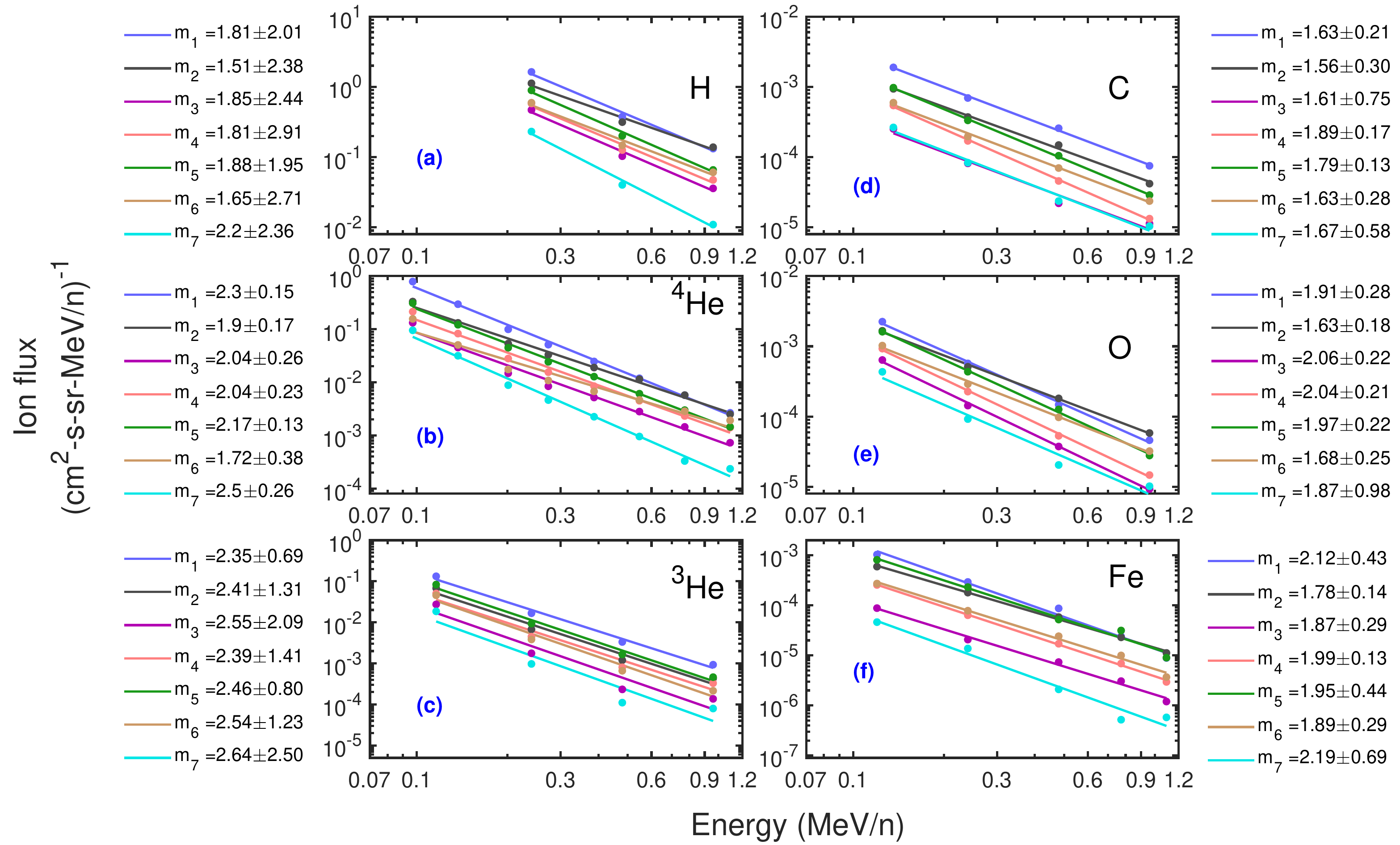}
\caption{Plots of two years’ ``quiet'' time averaged differential directional flux vs Energy for the six elements under consideration. Spectral slopes (m$_i$'s) of the fitted lines for different phases of solar cycles 23 and 24 are also shown in different colors. The margin of errors in the spectral index estimation within 95$\%$ confidence bounds are written on the left for H, $^4$He \& $^3$He and on the right for C, O $\&$ Fe respectively.\label{fig:fig6}}
\end{figure}

In order to evaluate the variabilities of spectral indices of different suprathermal elements during different phases of SC23 and SC24, seven phases are identified from the SSN data spanning from March 1998 to August 2020. The duration of each phase is two years. These phases are - (1) maximum of SC23 (from 11 November 1999 to 11 November 2001), (2) descending phase of SC23 (from 15 July 2003 to 15 July 2005), (3) minimum of SC23-24 (from 05 July 2007 to 05 July 2009), (4) ascending phase of SC24 (from 24 September 2009 to 24 September 2011), (5) maximum of SC24 (from 20 April 2012 to 20 April 2014), (6) descending phase of SC24 (from 08 July 2015 to 08 July 2017) and (7) minimum of SC24-25 (from 31 August 2018 to 31 August 2020). ``Quiet'' time suprathermal ion fluxes of each element are then averaged over each phase and plotted against corresponding energies (represented by colored dots) as shown in Figure \ref{fig:fig6}. Each subplot in Figure \ref{fig:fig6} corresponds to the spectra of an element mentioned at the right upper corner of each subplot. Lines with seven different colors are the least square fitted lines corresponding to the seven phases of solar cycles as mentioned earlier. The spectral indices denoted by m$_{i}$’s (where i = 1, 2, 3, 4, 5, 6 and 7) are the estimated mean slopes of the fitted lines corresponding to different phases. The margin of errors (MoE, which sets a lower and upper bound on the estimated value corresponding to a specified confidence level) within 95$\%$ confidence bounds in estimating the mean slopes are also mentioned here. It is to be noted that MoE depends on variability of the data points, sample size, and confidence level (e.g. \citealp{Agresti_1998}). Bootstrap sampling \citep{Tibshirani_1993} is one of the useful methods to calculate confidence interval and MoE of an estimated (fitted) parameter like spectral index.     

\begin{figure}[ht!]
\plotone{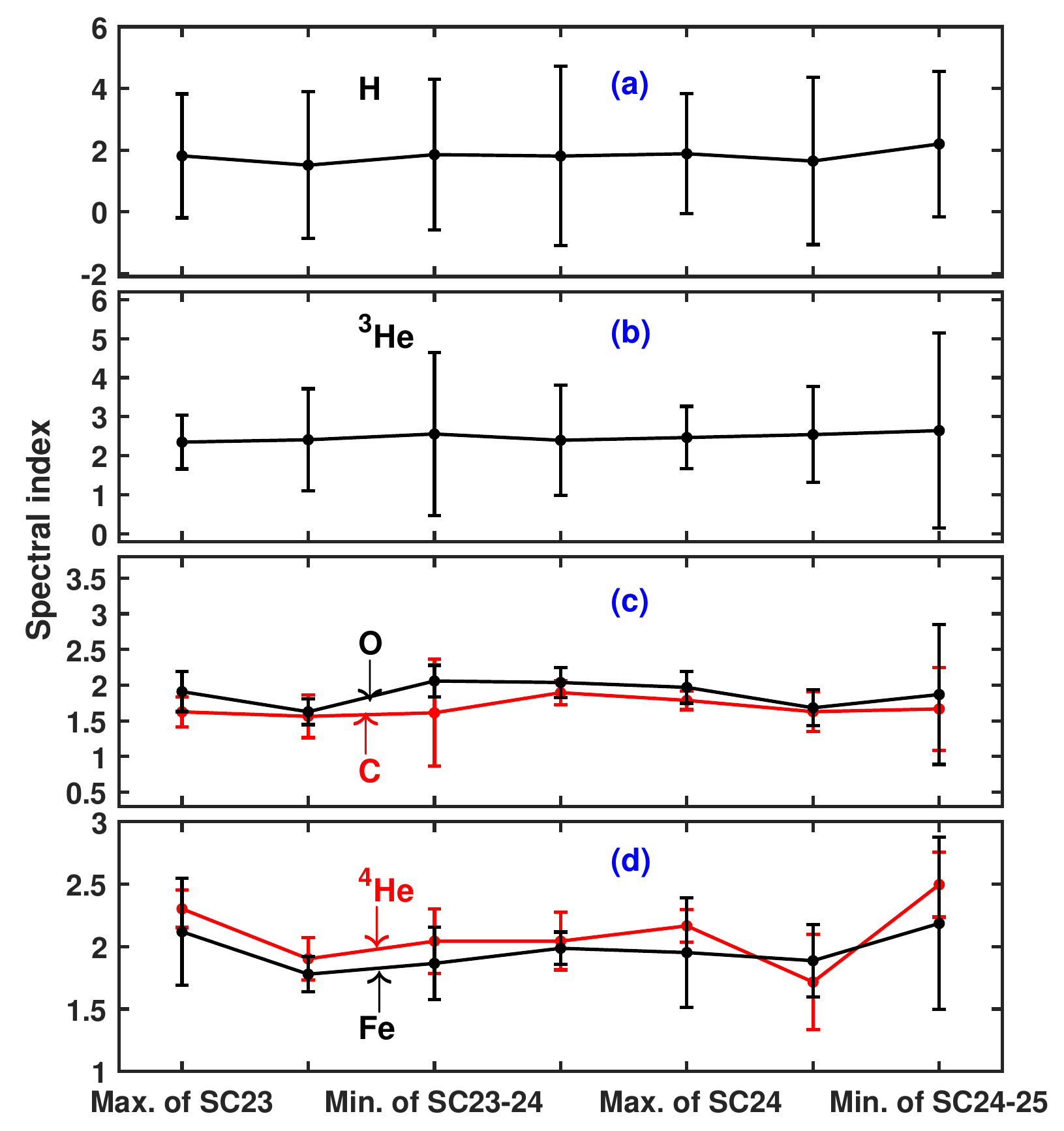}
\caption{Variations of spectral indices of (a) H, (b) $^3$He and comparative variations of (c) C (red) and O (black) and (d) $^4$He (red) and Fe (black) with respect to different phases of solar cycles starting from the maximum of SC23 up to the minimum of SC24-25. Margin of errors corresponding to 95$\%$ confidence interval, in estimating the spectral indices as shown in Figure \ref{fig:fig6}, are set as error bars..\label{fig:fig7}}
\end{figure}

Figure \ref{fig:fig7} summarizes the results shown in Figure \ref{fig:fig6}. It shows the variations of spectral indices of (a) H, (b) $^3$He, (c) C and O, and (d) $^4$He and Fe with different phases of solar cycles as mentioned above. The MoEs are also shown. Note, the Y-axis scales are made different for each subplot in Figure \ref{fig:fig7} for better visualization purpose.

Figure \ref{fig:fig7}(a) shows the variations in mean spectral indices for suprathermal H at different phases of solar cycles. The mean spectral index is found to vary from 1.51 (mean m$_2$) to 1.88 (mean m$_5$) except for that at the minimum of SC24-25 25 when it changes to 2.2 (mean m$_7$). It can be seen that the spectral index changed significantly during the minimum of SC24-25. However, since the variability of fluxes are large and number of data points is only three (corresponding to three energy channels), the MoE is relatively large. This argument is justified as Bootstrap sampling depends on the number of original data points and the more the data points, the better is the normal approximation used in Wald's method (see for example, \citealp{Agresti_1998}). Therefore, larger sample size and smaller variability of the sample data lead to smaller MoE. 

Figure \ref{fig:fig7}(b) shows the corresponding variation of spectral indices of $^3$He similar to what is shown in Figure \ref{fig:fig7}(a). It is noted that m$_i$'s of $^3$He vary from 2.35 (mean m$_1$) to 2.64 (mean m$_7$). The MoEs are also quite high (except for MoEs of mean m$_1$ (0.69) and mean m$_6$ (0.80)). Comparative variations of spectral indices of C and O (Figure \ref{fig:fig7}(c)) reveal that these two elements go almost hand in hand over all the phases of solar cycles with the largest difference between mean m$_i$'s appears to occur during the minimum of SC23-24. This is in contrast to $^4$He and Fe wherein the largest difference between the spectral slopes seems to occur during the minimum of SC24-25.  Spectral indices (from m$_1$ to m$_5$) of $^4$He and Fe vary almost hand in hand in between maximum of SC23 and maximum of SC24. However, during the descending phase of SC24, m$_6$ of $^4$He reaches the minimum value (1.72) and in the next phase, the index m$_7$ of $^4$He changes abruptly to the maximum value (2.50). Such abrupt change is not seen in case of Fe during the minimum of SC24-25. 

\section{Discussion}

Variations of ``quiet''-time suprathermal heavy ion ratios (C/O, Fe/O and $^3$He /$^4$He) with solar activity were reported in the past (e.g. \citealp{Dayeh_et_al_2009, Dayeh_et_al_2017, Kecskemety_et_al_2011} etc.). It was also suggested that these ratios exhibit SEP-like signature during solar maximum and CIR/solar wind like signature during solar minimum. Therefore, the contribution from SEP and CIR events to the suprathermal ion pool in the IP medium during solar maximum and minimum respectively have been indicated in the past. Our investigation explicitly brings out not only the solar activity dependence (based on significant correlation coefficients) but also the time delays between SSN and ``quiet'' time suprathermal fluxes. Therefore, our results augment and consolidate the earlier results (e.g. \citealp{Dayeh_et_al_2017}) on the solar activity dependence of suprathermal particles in the IP medium. It appears that as the occurrence of SEP (both impulsive and gradual events) and CIR events vary with the solar activity, so does the suprathermal population in the IP medium. The most striking point in this context emerges when we see conspicuous lags on many occasions despite removal of transient events. This strongly suggests that the so-called ``quiet'' time suprathermal population in the interplanetary (IP) medium may possibly be composed of leftover particles from previous solar and IP transient events. Therefore, ``quiet'' time population of suprathermal population may not be truly ``quiet'' (even after removal of the transient events).
\par
This investigation, for the first time, shows that there exists negative lags between suprathermal $^4$He fluxes and SSN in SC24. This essentially means that suprathermal $^4$He flux levels started rising in the IP medium before the SSN started rising during SC24. This indicates towards source processes that dominate during the minimum phase of solar cycle. Since CIRs are the major source of energetic particles in the IP medium during the minimum of solar cycles, the negative lags of suprathermal $^4$He during the minimum of SC23-24 probably indicate towards the production of these suprathermal particles from the CIR events in the interplanetary medium. The deep minimum of SC23-34 provides an ideal background condition for this. This argument is supported by the negative lags observed in the 0.96 MeV/n $^3$He flux in SC24. It is known that impulsive flare events contribute to the energetic $^3$He population in the IP medium (e.g. \citealp{Mason_et_al_2002}. Therefore, negative lag in $^3$He fluxes may additionally suggest the role of the remnant flare particles that were accelerated by the CIR events during the minimum of solar cycle and contributed to the suprathermal ion pool.
\par
\cite{Mason_et_al_2012} used the ACE/ULEIS data from 1998 to 2011 and found that CIR event-averaged suprathermal Fe/O lagged behind SSN in SC23 but became more in phase in SC24. The results of \cite{Mason_et_al_2012} also got extended and supported by the work of \cite{Allen_et_al_2019} wherein similar behavior for the CIR averaged suprathermal Fe/O (in the energy range of 0.32-0.45 MeV/n) is brought out using ACE/ULEIS data from both SC23 and 24 (1998-2018). Interestingly, the present results that deal with ``quiet'' time suprathermal populations, show consistency with the results obtained by both \cite{Mason_et_al_2012} and \cite{Allen_et_al_2019} as far as the difference in lags of the suprathermal Fe is concerned in both the cycles. This result also indicates towards the contribution of CIR-associated particles in the ``quiet'' time suprathermal particles. 
\par
Drastic changes in the spectral index of $^4$He occur after the maximum of SC24. In the descending phase of SC24, mean spectral index (m$_6$) of $^4$He reaches its minimum value at 1.72. In the next phase, the index (m$_7$=2.50) is the highest among all the phases. The change is greater than any of the margin of errors for $^4$He. On the other hand, spectral indices of Fe vary quite smoothly and reaches the maximum in the minimum of SC24-25. These variations in the spectral indices of $^4$He and Fe suggest that there were some differences in the production and (or) processing of these two suprathermal elements in the IP medium. Earlier, we have noted different lags of these two elements in SC23 and SC24. Therefore, it is clear that as far as the generation and processing of suprathermal $^4$He and Fe are concerned, there exists significant differences in SC24. In fact, SC24 is special in many ways. One such specialty is that SC24 is the weakest solar cycle in the last century and succeeds an extended minimum. \cite{Janardhan_et_al_2018} also reported unusual polar field reversal in the Sun. Therefore, it is natural to assume that the differential behavior of suprathermal $^4$He and Fe in SC24 may be associated with the changes in the Sun itself. However, lags and spectral index variations of C and O look similar in both the solar cycles. This contradiction leads us to think about preferential processing of suprathermal particles in the IP medium. The acceleration of suprathermal population in the IP medium has also been suggested to depend on mass to charge (M/q) (e.g. \citealp{Drake_et_al_2009, Zhao_et_al_2017, Reames_2018}) and First Ionization Potential (FIP) effects (e.g. \citealp{Feldman_and_Widing_2002}). Owing to the significant M/q and FIP differences, it is possible that processing of $^4$He and Fe in the IP medium were affected in SC24. This argument gets credence from the fact that C and O have similar FIP and M/q. Hence, no differential changes are noticed in lags and spectral indices of C and O in SC23 and SC24. Therefore, the present investigation suggests towards sensitive dependence of the generation of suprathermal population in the IP medium on the FIP and M/q. In the context of SC24 being different, a few other results are also relevant here. \cite{Mewaldt_et_al_2015} revealed that the fluence of the SEP particles were conspicuously lower during the initial phase of SC24 and this reduction is attributed to two factors – a lower IMF in the IP medium and depletion in the seed population (suprathermal particles). While the reduced IMF reduces the efficiency of acceleration leading to under-population in the SEP energy domain (e.g. \citealp{Gopalswamy_et_al_2014, Mewaldt_et_al_2015, Allen_et_al_2019}), lower densities of seed population may limit the maximum energies that particles can be accelerated to by the Alfven waves through the wave-particle interaction at the shock front (e.g. \citealp{Li_and_Lee_2015}). Another possibility is the changes in the spectral slopes through variations in the pick-up ion populations in the IP medium. 
\par
The relative importance of the acceleration and deceleration processes in the changes of suprathermal $^4$He that we see during the minimum of SC24-25 also deserves attention at this stage. \cite{Dayeh_et_al_2017} discussed that suprathermal tails are either generated due to the (1) continuous acceleration of the seed populations in the IP medium (e.g.\citealp{Fisk_and_Gloeckler_2006, Fisk_and_Gloeckler_2008, Fisk_and_Gloeckler_2014, Zank_et_al_2014} etc.) or due to the (2) deceleration of energetic particles from the previous solar and IP events (e.g. \citealp{Fisk_and_Lee_1980, Giacalone_et_al_2002} etc.). Earlier studies have suggested that lower energy part of the CIR (or Stream Interface Region, SIR) associated suprathermal population gets affected primarily by local acceleration processes (e.g., \citealp{Schwadron_et_al_1996, Giacalone_et_al_2002, Ebert_et_al_2012, Filwett_et_al_2017, Filwett_et_al_2019, Allen_et_al_2020, Allen_et_al_2021}) while the higher energy part gets affected by the shock acceleration occurring far away (e.g., \citealp{Ebert_et_al_2012, Filwett_et_al_2019}). Therefore, importance of acceleration, whether local or shock induced, is undeniable in the observed changes that are reported in the present investigation. \cite{Schwadron_et_al_2010} theorized that variable superpositions of stochastic processes (distribution functions represented by exponential and Gaussian functions) could give rise to power law distribution function with exponent of -5 (f $\alpha$ v-5) (or -1.5 in differential intensity w.r.t. energy approach).  This work essentially suggests that variable acceleration and heating processes may be in operation in IP medium for the generation of suprathermal particles. \cite{Antecki_et_al_2013} adopted stochastic acceleration under pressure balance condition - time scale of acceleration is in balance with timescale of adiabatic cooling in the solar wind as an effective mean of producing -1.5 spectral index. On the other hand, possible role of deceleration processes for the observed changes in the suprathermal $^4$He particles during the minimum of SC24-25 cannot be ruled out also. This is because the particles that are accelerated at the distant shock fronts can be thrown back into the inner heliosphere leading to increased scattering and magnetic cooling processes. This may result in modulation in intensity and spectral index (e.g. \citealp{Fisk_and_Lee_1980, Mason_et_al_1999, Zhao_et_al_2016, Allen_et_al_2021}). Most importantly, particles accelerated through previous transient events, while speeding through the IP medium, experience adiabatic expansion of the solar wind resulting into deceleration of these particles. As a consequence, the energetic particles slow down and enter into the suprathermal pool. According to \cite{Fisk_and_Lee_1980}, this type of particles is expected to show a rollover below $\sim$ 0.5 MeV/n in the spectra. However, no such rollover is observed by \cite{Mason_et_al_1997}. We also don’t see such rollover feature in the present investigation. Nevertheless, although difficult to comment, the relative role of deceleration processes on the significant changes in the spectral slope of $^4$He during SC24-25 cannot be ruled out. It seems, therefore, reasonable to argue that variable contributions of acceleration and deceleration or distribution functions may lead to variation in spectral indices with the varying solar activity. These processes might have undergone changes in the minimum of SC24-25 so as to cause changes in the spectral indices as far as heavier ions like Fe and $^4$He are concerned. These aspects need more critical attention in future and the multi-directional suprathermal particle measurements on-board India’s forthcoming Aditya-L1 mission \citep{Goyal_et_al_2018} may throw important light on some of these issues. 

\section{Summary}

In this work, by analyzing 22 years’ of flux variations of suprathermal H, $^3$He, $^4$He, C, O and Fe measured by ACE spacecraft from the first Lagrangian point of the Sun-Earth system, we show that these particles follow solar cycle variation with varying lags or time delays. These time delays are shown to vary with energy and element. It is inferred that suprathermal population in the IP medium during ``quiet'' times are not free from particles originated from earlier transient events. Further, this investigation, for the first time, reveals that suprathermal Fe and $^4$He show discernible changes in lags and spectral slopes in solar cycle 24. It is hypothesized that these changes in lags and spectral slopes may result from acceleration/deceleration processes that depend on mass-by-charge and first ionization potential. This suggestion gets credence from the behavior of C and O wherein no such differential changes are noticed. Further investigations are needed to understand the underlying physical processes that are responsible for the differential changes in suprathermal helium and iron in the interplanetary medium in solar cycle 24.

\section{Acknowledgements}
We are grateful to the Principal Investigator (PI) and all the members of the ULEIS team for constructing the ULEIS instrument and thereafter, generating and managing the dataset used in this work. We also thank WDC-SILSO, Royal Observatory of Belgium, Brussels for the daily estimated sunspot number data used in this paper. We express our sincere gratitude to the Department of Space, Government of India for supporting this work.

\section{Data sources}
The suprathermal particle flux data used in this paper are available at \url{https://cdaweb.gsfc.nasa.gov/cgi-bin/eval2.cgi}. The daily estimated sunspot number data can be downloaded from \url{ http://www.sidc.be/silso/datafiles}.

\bibliography{Bijoy_paper_updated}{}
\bibliographystyle{aasjournal}
\appendix

\section{Supplementary materials}
\begin{figure}[ht!]
\plotone{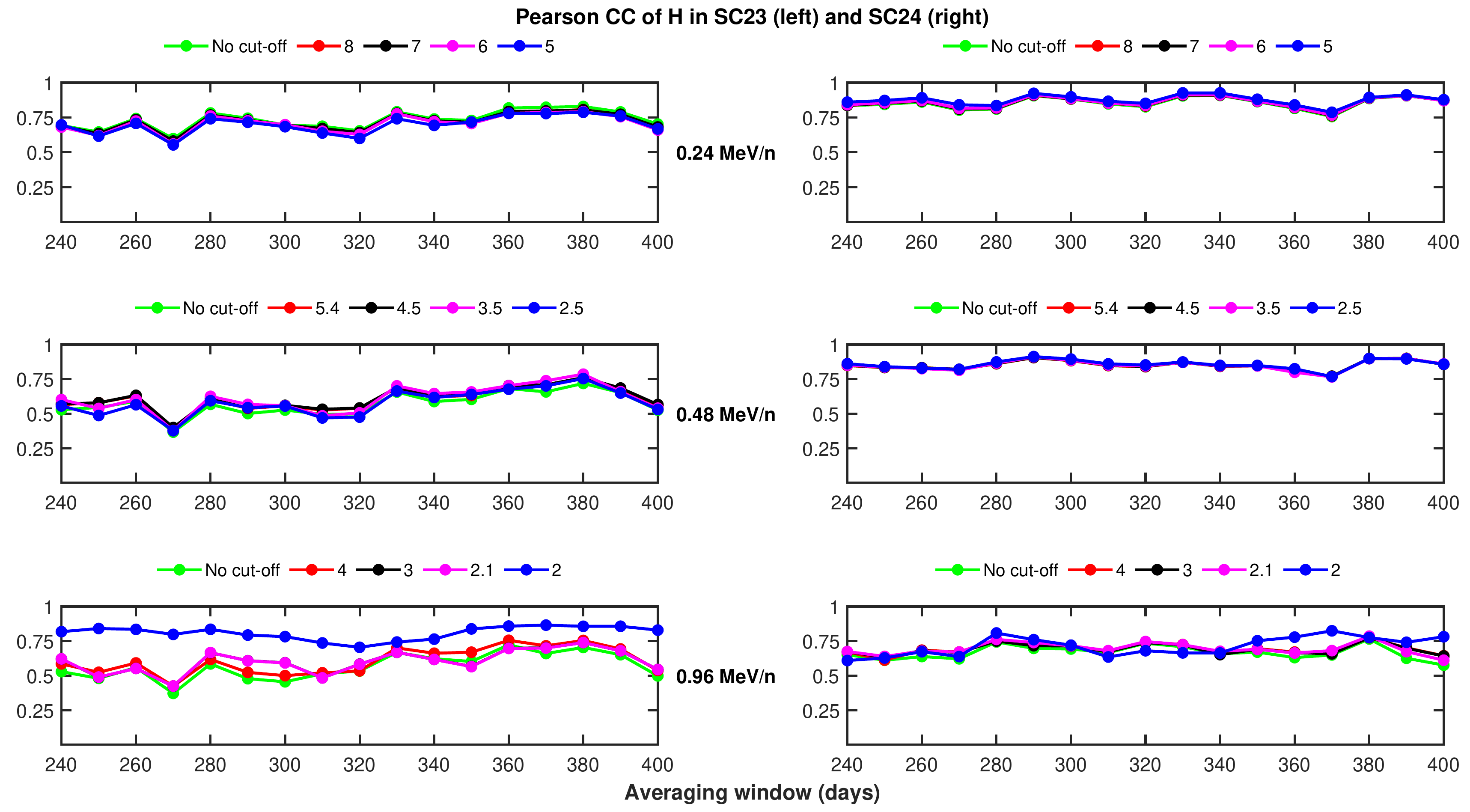}
\renewcommand{\thefigure}{S1}
\caption{Variations of Pearson correlation coefficients (CCs) between averaged and shifted H fluxes at different energies (written in the middle column) and averaged sunspot numbers (SSN) with respect to the averaging window during SC23 (left) and SC24 (right) are shown. Fluxes are shifted based on the calculated lags with respect to the SSN variation. Green dots are Pearson CCs when there is no cut-off on the fluxes (as shown in Figure 3 in the main script). Red, black, pink, and blue dots represent the Pearson CCs obtained in the similar way after putting cut-offs on the fluxes. Threshold values of fluxes in units of particles/(cm2-s-sr-MeV/n) are written  above each subplot.\label{fig:S1}}
\end{figure}

\begin{figure}[ht!]
\plotone{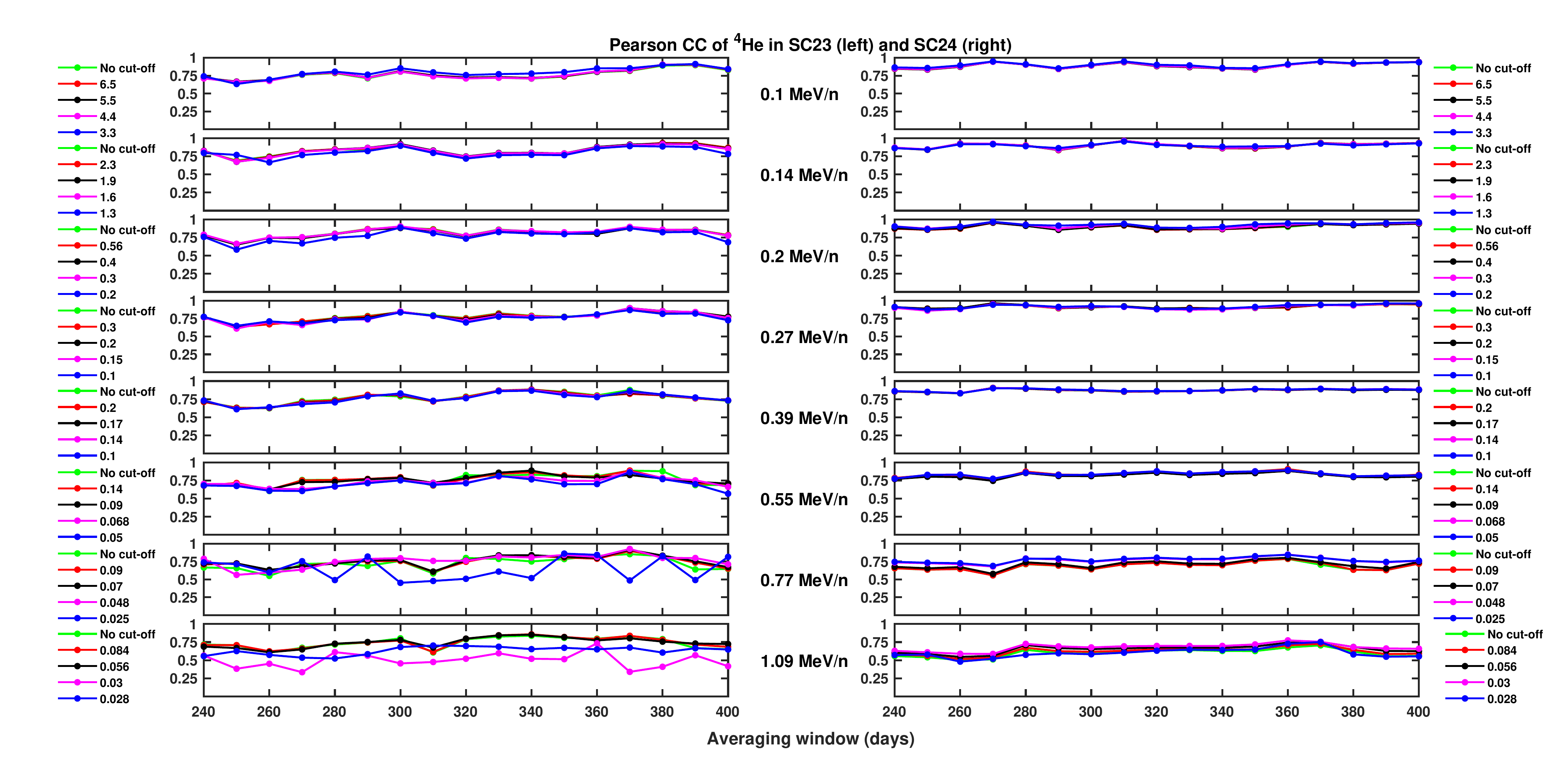}
\renewcommand{\thefigure}{S2}
\caption{Variations of Pearson correlation coefficients (CCs) between averaged and shifted $^4$He fluxes at different energies (written in the middle column) and averaged SSN with respect to the averaging window during SC23 (left) and SC24 (right) are shown. Fluxes are shifted based on the calculated lags with respect to the SSN variation. Green dots are Pearson CCs when there is no cut-off on the fluxes (as shown in Figure 3 in the main script). Red, black, pink, and blue dots represent the Pearson CCs obtained in the similar way after putting cut-offs on the fluxes. Threshold values of fluxes in units of particles/(cm2-s-sr-MeV/n) are written  on the left and right of each pair of subplots corresponding to SC23 and SC24 respectively.\label{fig:S2}}
\end{figure}

\begin{figure}[ht!]
\plotone{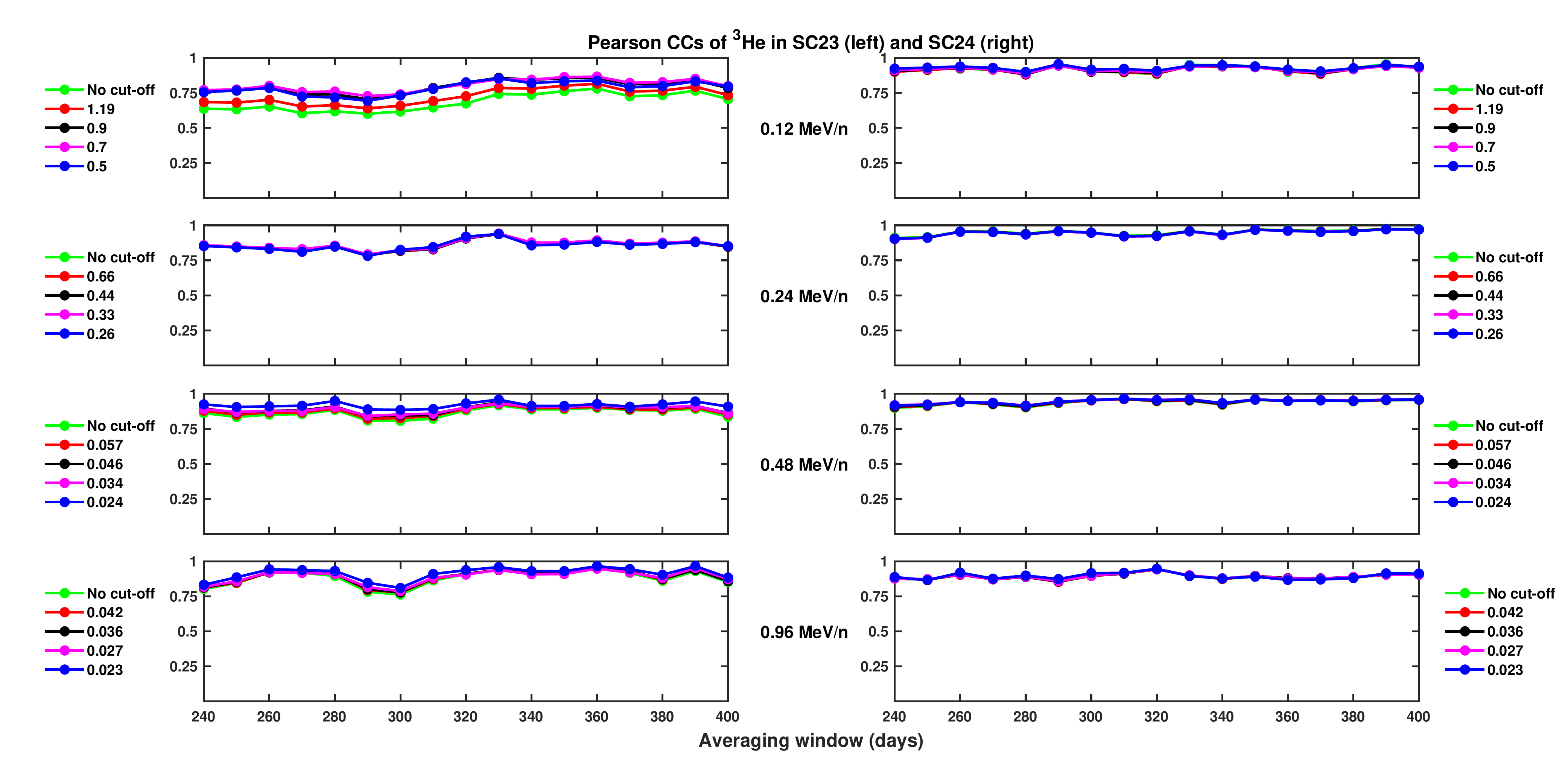}
\renewcommand{\thefigure}{S3}
\caption{Similar to Figure \ref{fig:S2} but for $^3$He.\label{fig:S3}}
\end{figure}

\begin{figure}[ht!]
\plotone{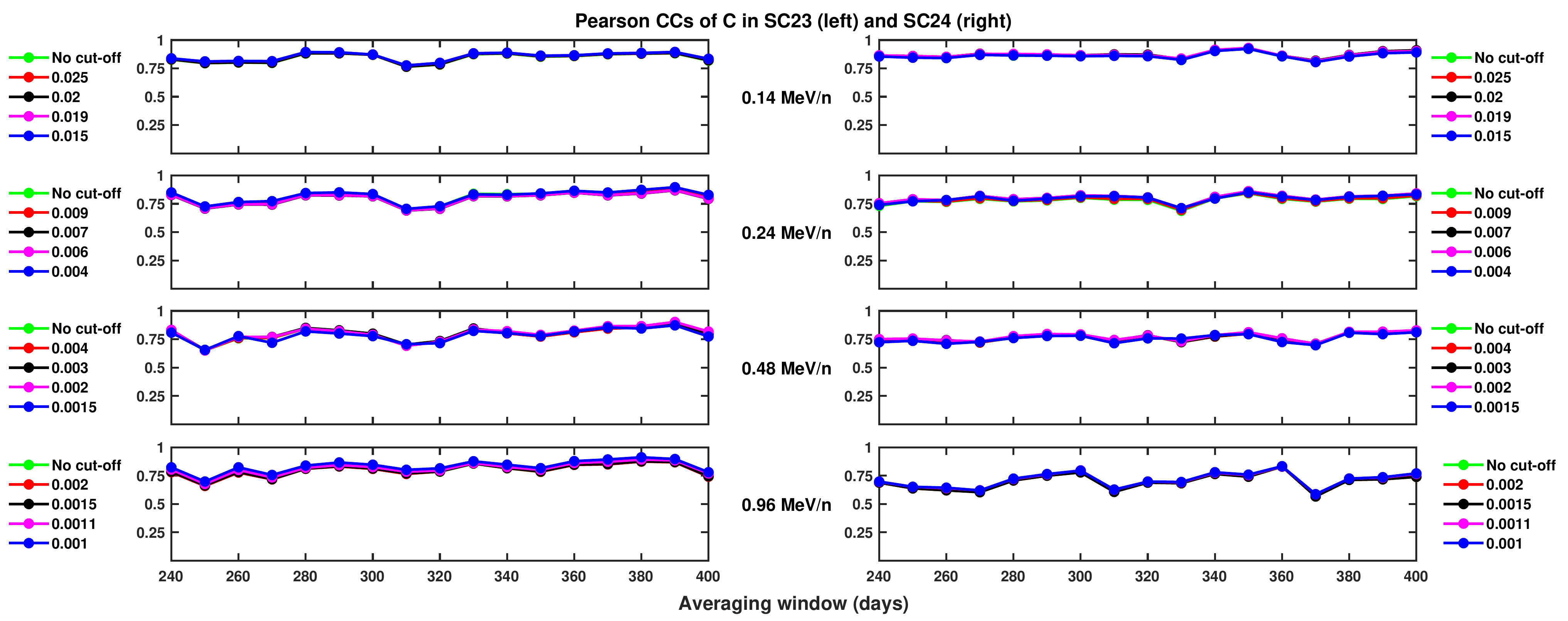}
\renewcommand{\thefigure}{S4}
\caption{Similar to Figure \ref{fig:S2} but for C.\label{fig:S4}}
\end{figure}

\begin{figure}[ht!]
\plotone{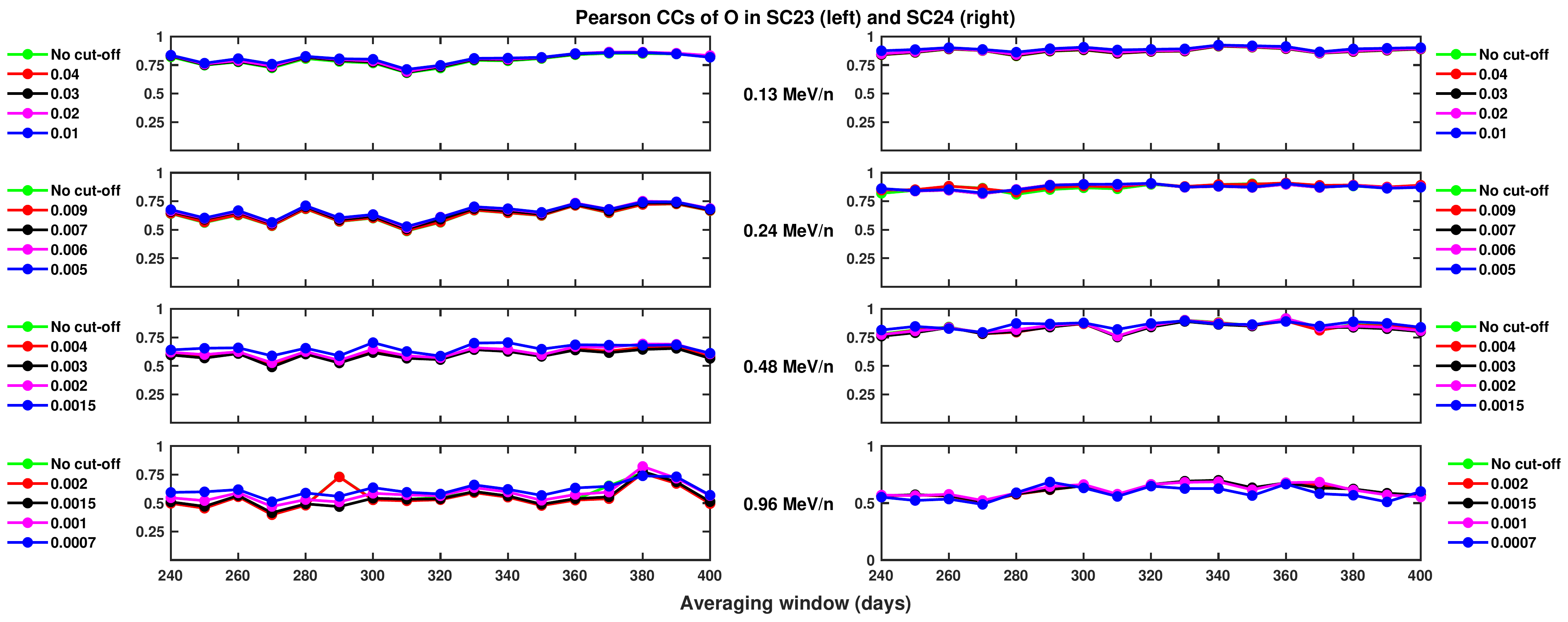}
\renewcommand{\thefigure}{S5}
\caption{Similar to Figure \ref{fig:S2} but for O.\label{fig:S5}}
\end{figure}

\begin{figure}[ht!]
\plotone{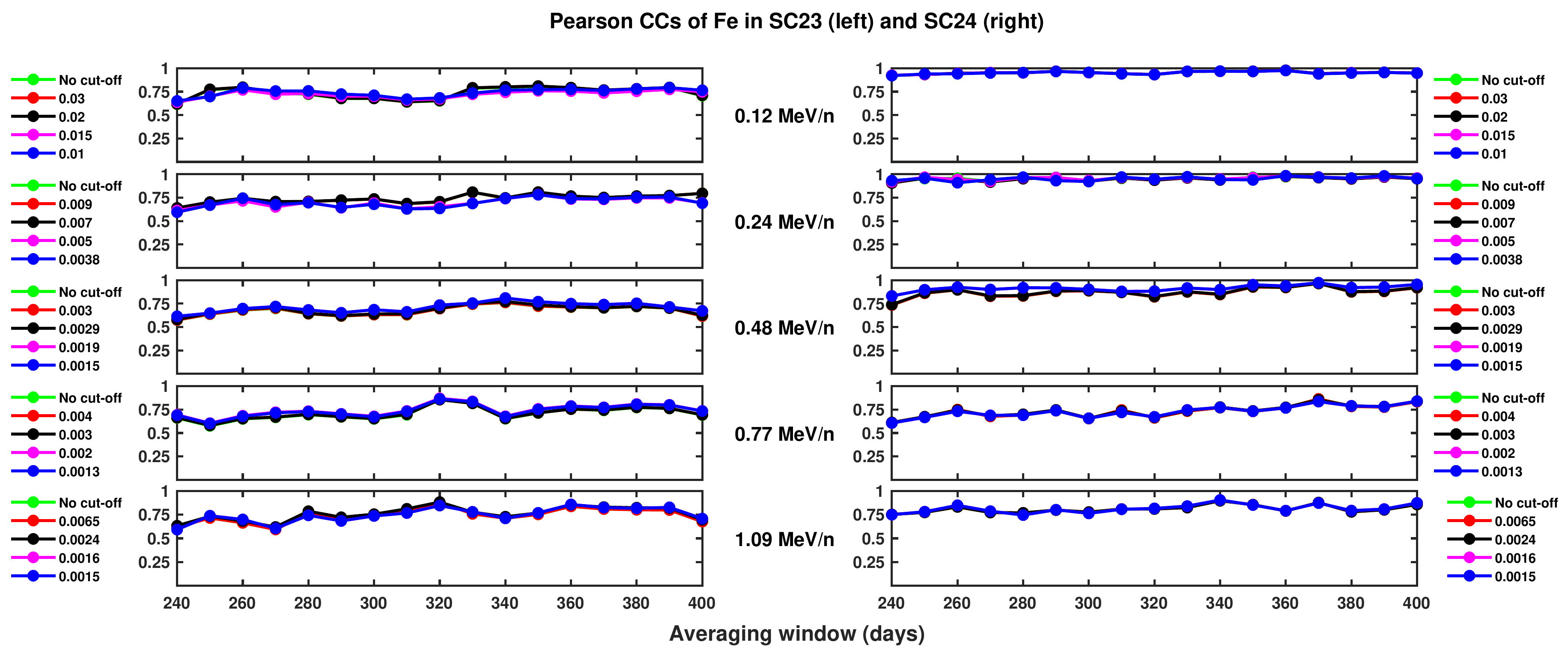}
\renewcommand{\thefigure}{S6}
\caption{Similar to Figure \ref{fig:S2} but for Fe.\label{fig:S6}}
\end{figure}

\begin{figure}[ht!]
\plotone{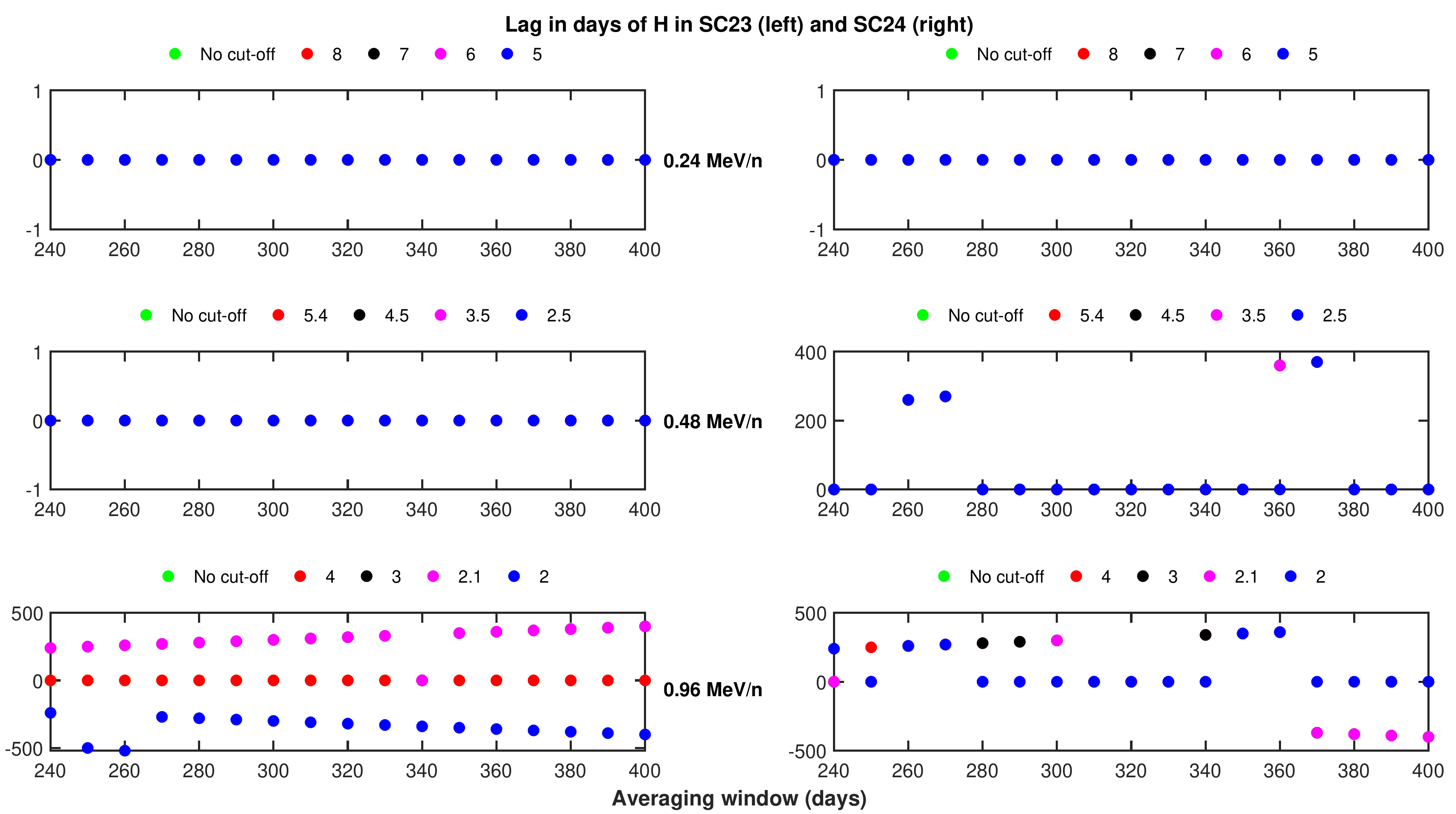}
\renewcommand{\thefigure}{S7}
\caption{Variations of lags between averaged and shifted H fluxes at different energies (mentioned in between the sub-plots) and averaged sunspot numbers (SSN) with respect to the averaging window during SC23 (left) and SC24 (right) are shown. Fluxes are shifted based on the calculated lags with respect to the SSN variation. Green dots are lags when there is no cut-off on the fluxes (as shown in Figure 4 in the main script). Red, black, pink, and blue dots represent the Pearson CCs obtained in the similar way after putting cut-offs on the fluxes. Threshold values of fluxes in units of particles/(cm2-s-sr-MeV/n) are written  above each subplot.\label{fig:S7}}
\end{figure}

\begin{figure}[ht!]
\plotone{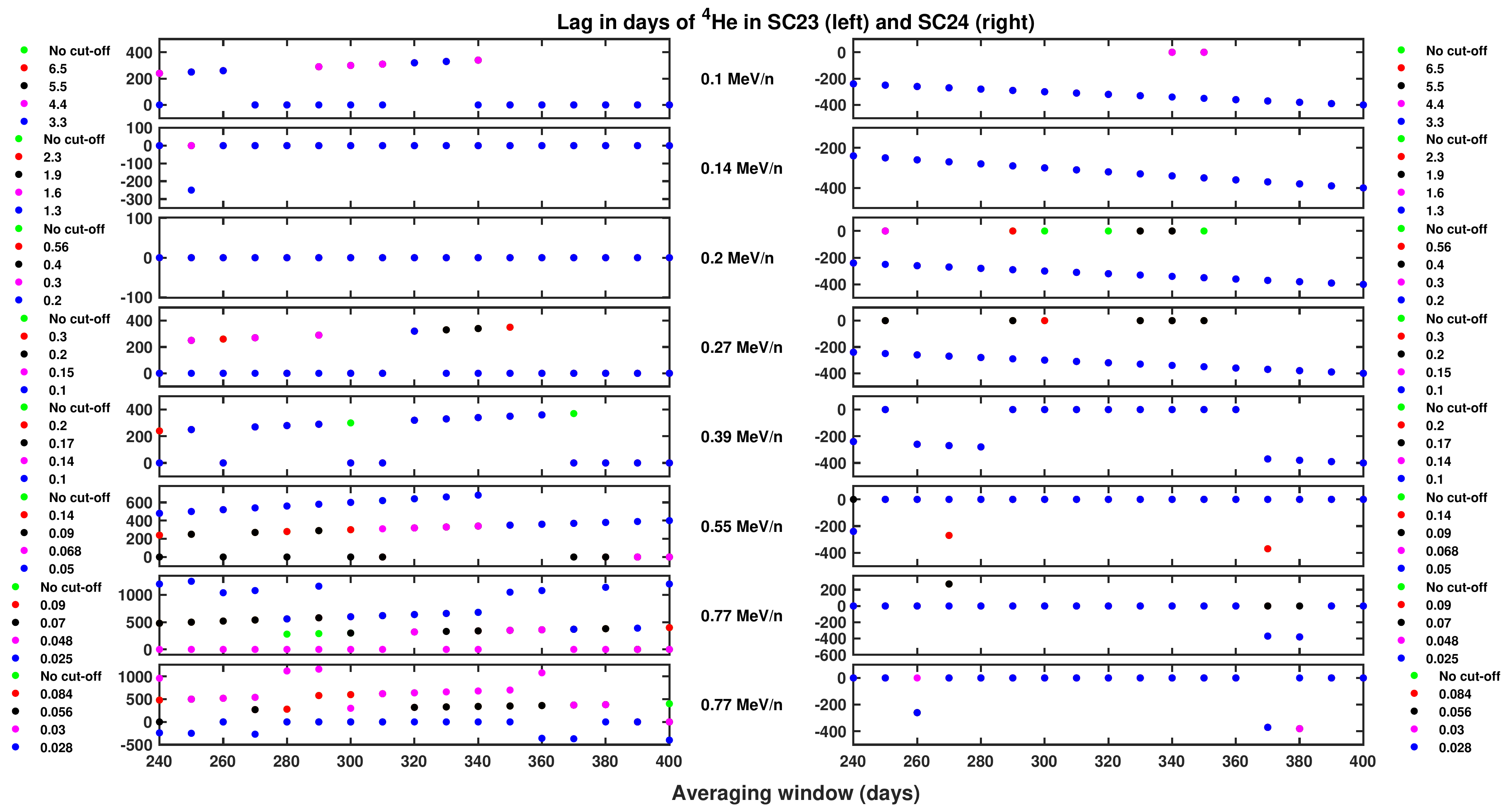}
\renewcommand{\thefigure}{S8}
\caption{Similar as Figure \ref{fig:S7} but for $^4$He. Also, the legends are shown adjacent to the sub-plots.\label{fig:S8}}
\end{figure}

\begin{figure}[ht!]
\plotone{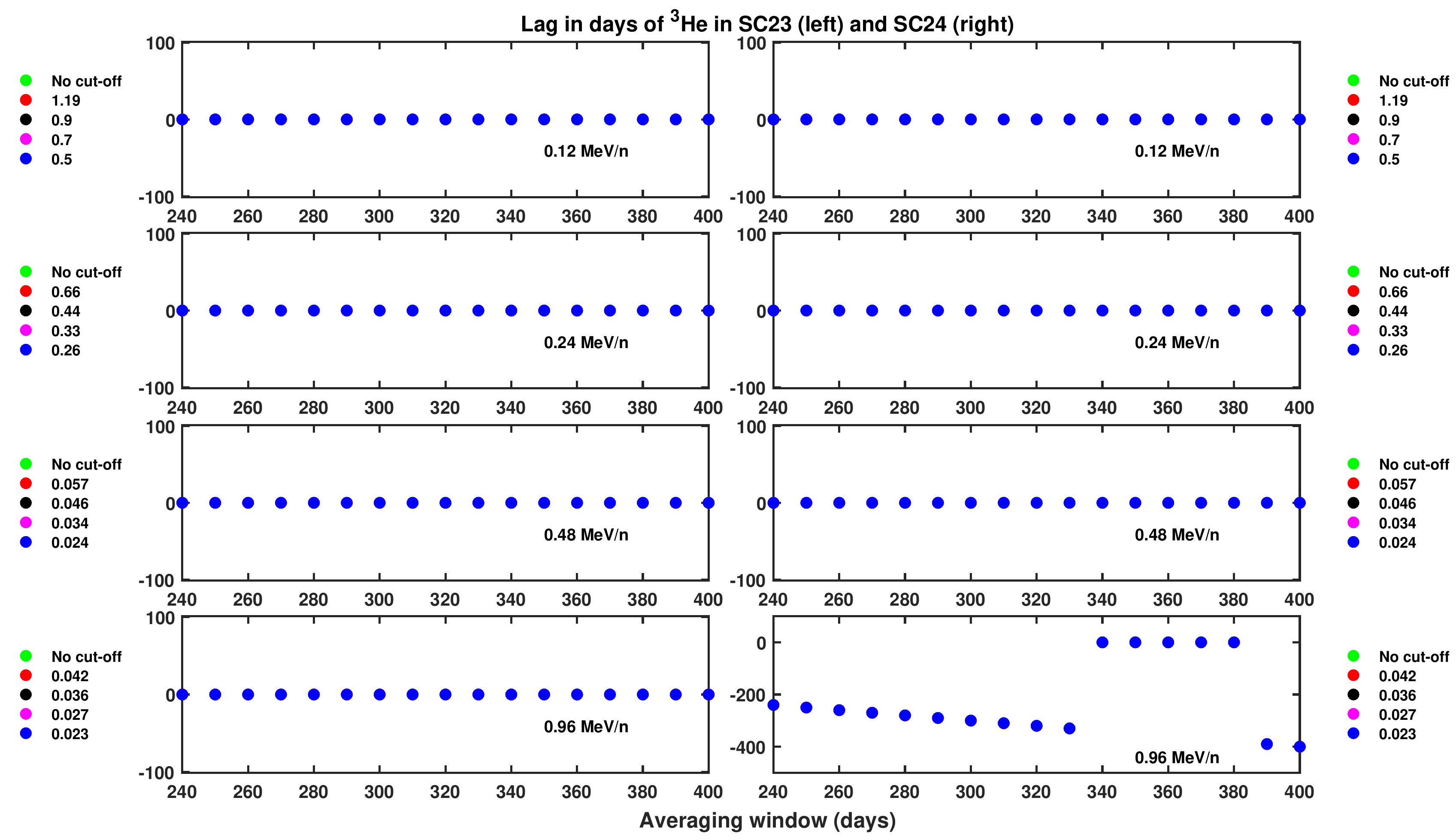}
\renewcommand{\thefigure}{S9}
\caption{Similar as Figure \ref{fig:S8} but for $^3$He. Energy channels are mentioned inside the sub-plots.\label{fig:S9}}
\end{figure}

\begin{figure}[ht!]
\plotone{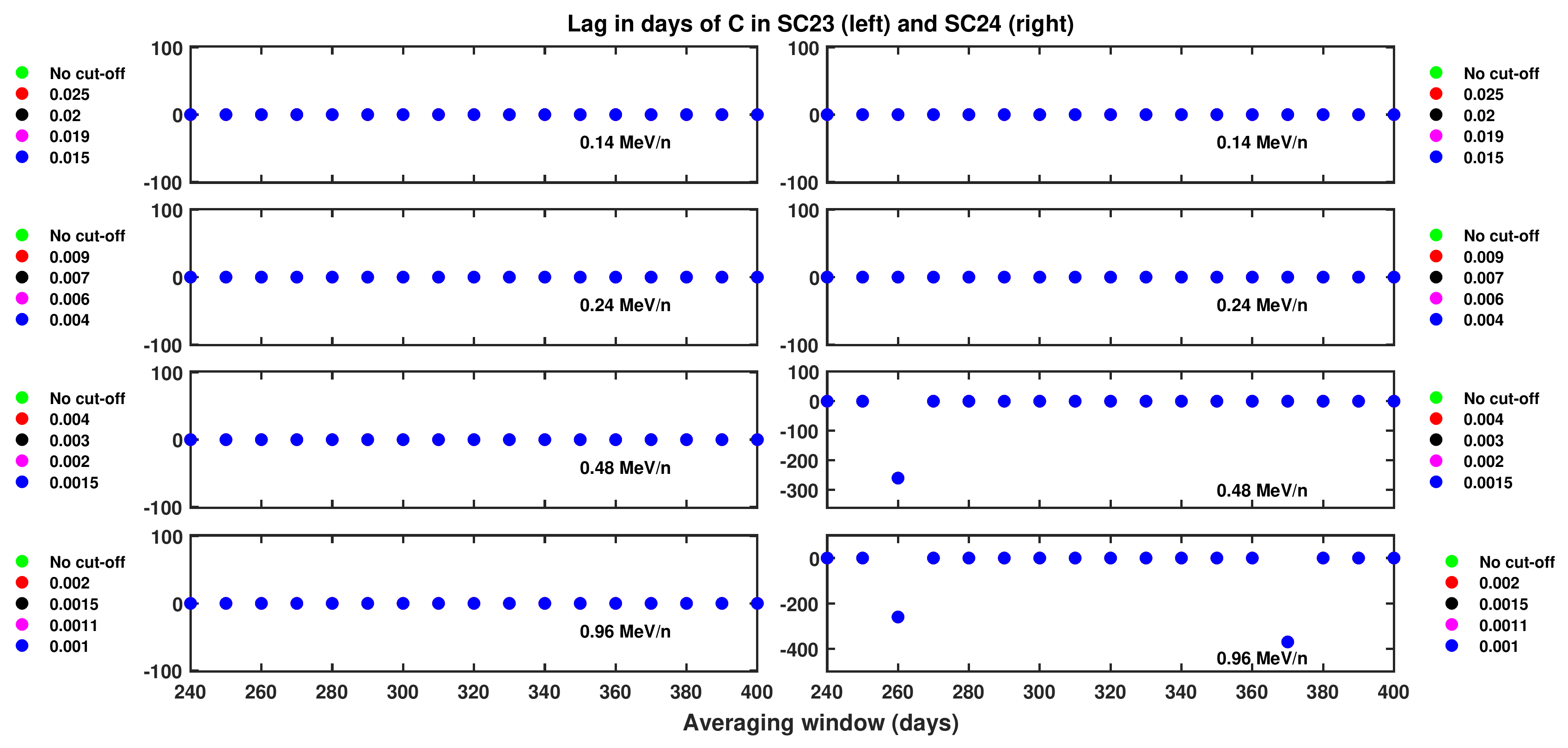}
\renewcommand{\thefigure}{S10}
\caption{Similar as Figure \ref{fig:S9} but for C.\label{fig:S10}}
\end{figure}

\begin{figure}[ht!]
\plotone{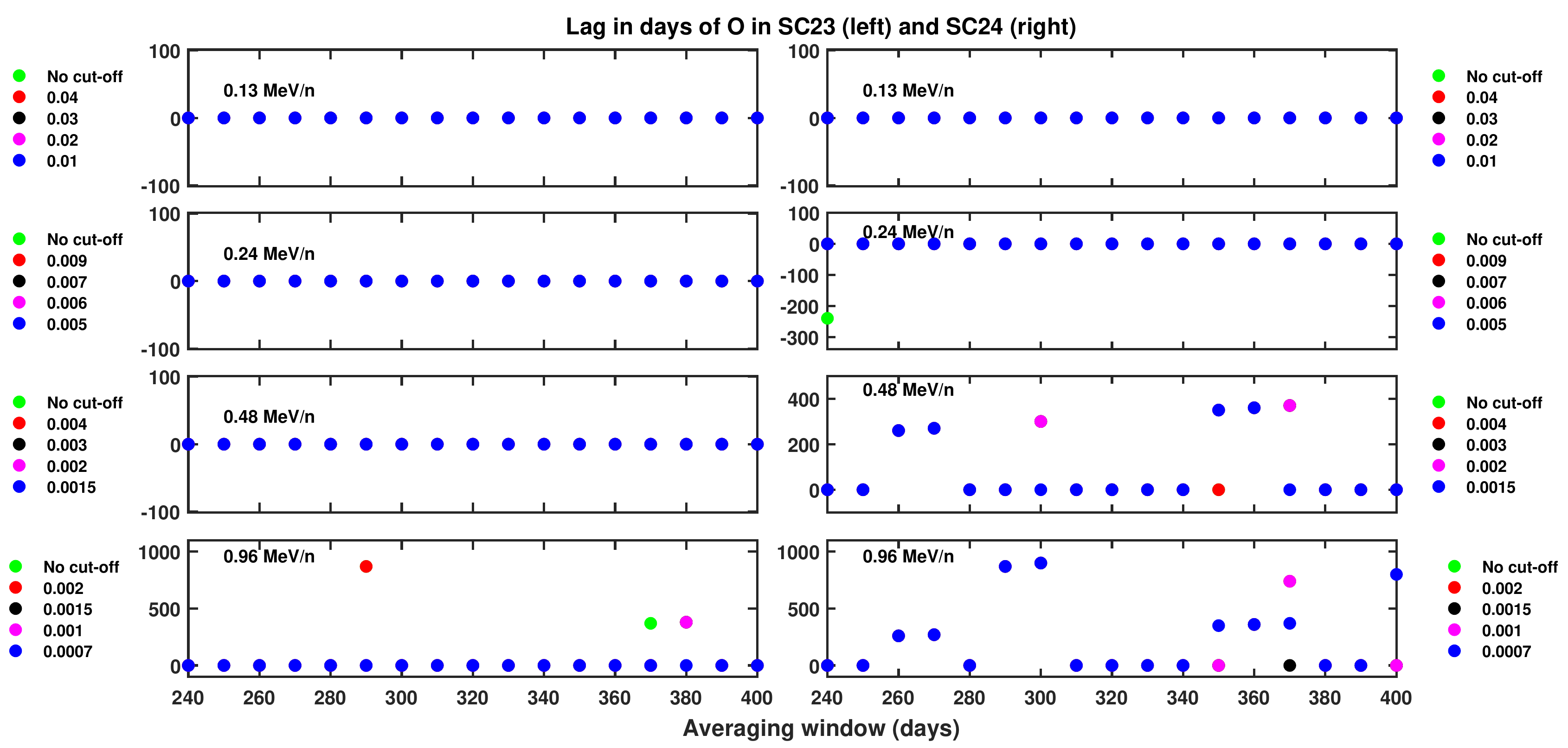}
\renewcommand{\thefigure}{S11}
\caption{Similar as Figure \ref{fig:S9} but for O.\label{fig:S11}}
\end{figure}

\begin{figure}[ht!]
\plotone{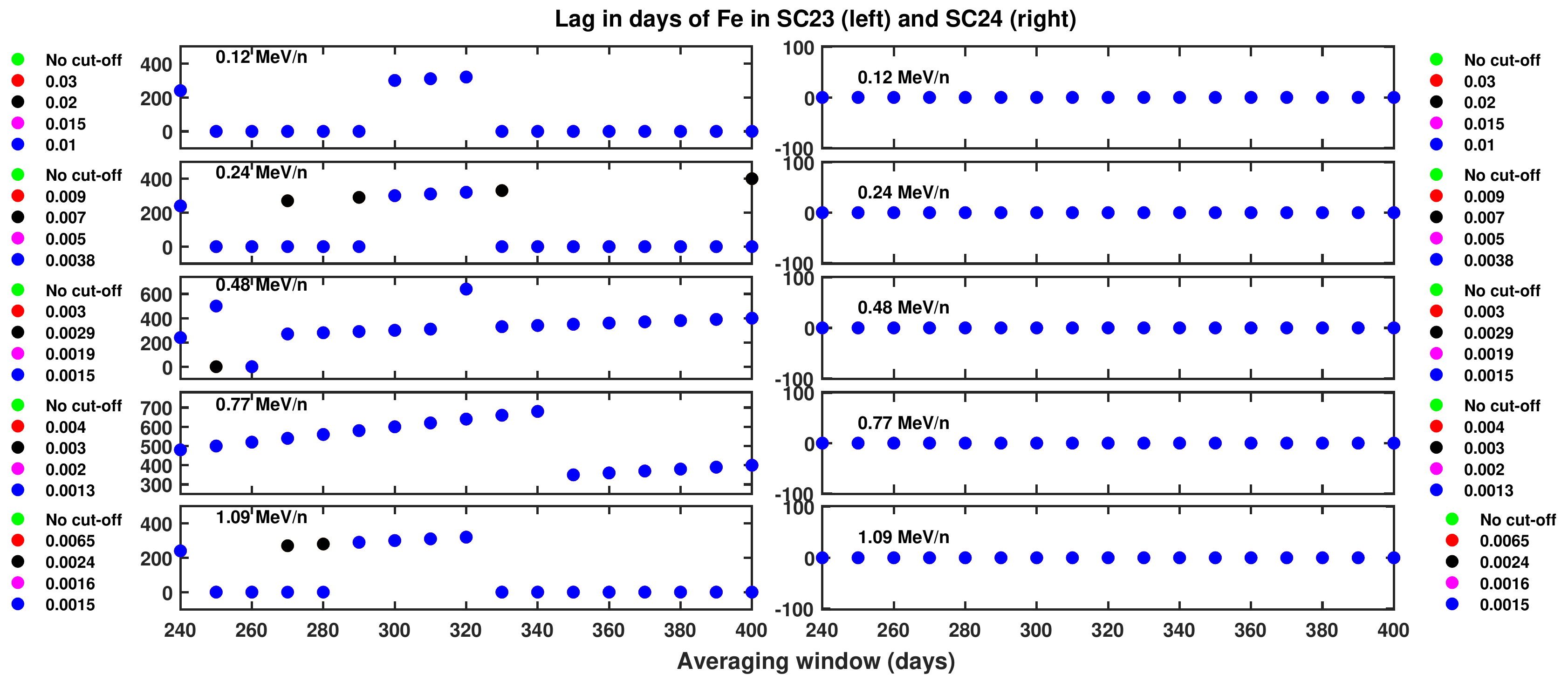}
\renewcommand{\thefigure}{S12}
\caption{Similar as Figure \ref{fig:S9} but for Fe.\label{fig:S12}}
\end{figure}

Rigorous sensitivity tests to validate the results regarding lags of suprathermal particles with respect to SSN variation are performed. These plots reveal how the correlation between quiet suprathermal fluxes and SSN varies depending on flux thresholds. Four sets of flux thresholds have been selected at random in decreasing order for each energy channel of each element. We have repeated the same exercise depicted in the main text to calculate the lags and Pearson correlation coefficients between SSN and fluxes with the above mentioned cut-offs. The results of the same are shown in Figure \ref{fig:S1}-\ref{fig:S12}. Figure \ref{fig:S1}-\ref{fig:S6} show the variations in Pearson correlation coefficients for H, $^4$He, $^3$He, C, O, and Fe respectively. It can be seen from Figure \ref{fig:S1} to Figure \ref{fig:S6} that irrespective of the thresholds on the flux, Pearson correlation coefficients do not change significantly if we compare these values with the CCs calculated by only removing transient events. Not only the CCs, the time delays between fluxes and SSN also don’t change depending on flux baselines (see Figure \ref{fig:S7}-\ref{fig:S12}). These give a direct credence to the methodology adopted here for generating quiet time data by removing the transient events.

\end{document}